\documentclass[prd,aps,
nofootinbib,floatfix,
preprintnumbers,
superscriptaddress]{revtex4}
\usepackage{graphicx}
\usepackage{epsfig}
\usepackage{rotating}
\usepackage{amssymb}
\usepackage{subfigure}
\usepackage{enumitem}
\usepackage{dsfont}
\usepackage{psfrag}
\usepackage{amsmath,euscript,array,mathrsfs}
\usepackage{slashed}
\usepackage{array}
\usepackage{youngtab}
\usepackage{color}
\usepackage{bbold}
\usepackage{subfiles}
  \usepackage{orcidlink}
\usepackage{transparent}
\usepackage{xcolor}
\usepackage{soul}
\usepackage{hyperref}
\hypersetup{
colorlinks = true,
linkcolor = blue,
linktocpage = true,
citecolor = blue
}

\newcommand{\beq}{\begin{equation}}
\newcommand{\eeq}{\end{equation}}
\newcommand{\beqs}{\begin{eqnarray}}
\newcommand{\eeqs}{\end{eqnarray}}
\newcommand{\Tr}{{\rm Tr}}

\newcommand{\be}{\begin{equation}}
\newcommand{\ee}{\end{equation}}
\newcommand{\ba}{\begin{array}}
\newcommand{\ea}{\end{array}}

\newcommand{\orcidauthorBENNETT}{0000-0002-1678-6701}
\newcommand{\orcidauthorLENZ}{0000-0001-5250-0005}
\newcommand{\orcidauthorLUCINI}{0000-0001-8974-8266}
\newcommand{\orcidauthorPIAI}{0000-0002-2251-0111} 
\newcommand{\orcidauthorFORZANO}{0000-0003-0985-8858}
\newcommand{\orcidauthorVADACCHINO}{0000-0002-5783-5602}
\newcommand{\orcidauthorHONG}{0000-0002-3923-4184}
\newcommand{\orcidauthorDELDEBBIO}{0000-0003-4246-3305}
\newcommand{\orcidauthorLUPO}{0000-0001-9661-7811}
\newcommand{\orcidauthorLIN}{0000-0003-3743-0840}
\newcommand{\orcidauthorLEE}{0000-0002-4616-2422}

\begin{document}

\title{Symplectic lattice gauge theories in the grid framework: Approaching the conformal window}

\author{Ed Bennett\,\orcidlink{\orcidauthorBENNETT}}
\email{E.J.Bennett@swansea.ac.uk}
\affiliation{Swansea Academy of Advanced Computing, Swansea University (Bay Campus), Fabian Way, SA1 8EN Swansea, Wales, United Kingdom}

\author{Peter A. Boyle}
\email{pboyle@bnl.gov}
\affiliation{School of Physics and Astronomy, University of Edinburgh, Edinburgh EH9 3FD, United Kingdom}
\affiliation{Physics Department, Brookhaven National Laboratory, Upton, NY 11973, USA}

\author{Luigi Del Debbio\,\orcidlink{\orcidauthorDELDEBBIO}}
\email{luigi.del.debbio@ed.ac.uk}
\affiliation{Higgs Centre for Theoretical Physics, School of Physics and Astronomy, 
The University of Edinburgh, Peter Guthrie Tait Road, Edinburgh EH9 3FD, UK}

\author{Niccolò Forzano\,\orcidlink{\orcidauthorFORZANO}}
\email{2227764@swansea.ac.uk}
\affiliation{Department of Physics, Faculty  of Science and Engineering,
Swansea University,
Singleton Park, SA2 8PP, Swansea, Wales, UK}

\author{Deog~Ki Hong\,\orcidlink{\orcidauthorHONG}}
 \email{dkhong@pusan.ac.kr}
\affiliation{
 Department of Physics, Pusan National University, Busan 46241, Korea
}%

\author{Jong-Wan Lee\,\orcidlink{\orcidauthorLEE}}
 \email{j.w.lee@ibs.re.kr}
\affiliation{
 Department of Physics, Pusan National University, Busan 46241, Korea
}%
\affiliation{
 Institute for Extreme Physics, Pusan National University, Busan 46241, Korea
}%
\affiliation{
Particle Theory  and Cosmology Group, Center for Theoretical Physics of the Universe, Institute for Basic Science (IBS), Daejeon, 34126, Korea
}

\author{Julian Lenz\,\orcidlink{\orcidauthorLENZ}}
\email{j.j.lenz@swansea.ac.uk}
\affiliation{Swansea Academy of Advanced Computing, Swansea University (Bay Campus), Fabian Way, SA1 8EN Swansea, Wales, United Kingdom}

\author{C.-J. David Lin\,\orcidlink{\orcidauthorLIN}}
 \email{dlin@nycu.edu.tw}
\affiliation{%
 Institute of Physics, National Yang Ming Chiao Tung University, 1001 Ta-Hsueh
Road, Hsinchu 30010, Taiwan
}%
\affiliation{
Centre for High Energy Physics, Chung-Yuan Christian University,
Chung-Li 32023, Taiwan
}%
\affiliation{
Centre for Theoretical and Computational Physics, National Yang Ming Chiao Tung University, 1001 Ta-Hsueh Road, Hsinchu 30010, Taiwan
}%

\author{Biagio~Lucini\,\orcidlink{\orcidauthorLUCINI}}
\email{B.Lucini@Swansea.ac.uk}
\affiliation{Swansea Academy of Advanced Computing, Swansea University (Bay Campus), Fabian Way, SA1 8EN Swansea, Wales, United Kingdom}
\affiliation{Department of Mathematics, Faculty of Science and Engineering, Swansea University (Bay Campus), Fabian Way, SA1 8EN Swansea, Wales, United Kingdom}

\author{Alessandro Lupo\,\orcidlink{\orcidauthorLUPO}}
\email{Alessandro.Lupo@ed.ac.uk}
\affiliation{Higgs Centre for Theoretical Physics, School of Physics and Astronomy, 
The University of Edinburgh, Peter Guthrie Tait Road, Edinburgh EH9 3FD, UK}

\author{Maurizio Piai\,\orcidlink{\orcidauthorPIAI}}
\email{m.piai@swansea.ac.uk}
\affiliation{Department of Physics, Faculty  of Science and Engineering,
Swansea University,
Singleton Park, SA2 8PP, Swansea, Wales, UK}

\author{Davide Vadacchino\,\orcidlink{\orcidauthorVADACCHINO}}
\email{davide.vadacchino@plymouth.ac.uk}
\affiliation{Centre for Mathematical Sciences, University of Plymouth, Plymouth, PL4 8AA, United Kingdom}

\definecolor{mygray}{gray}{0.9}
\colorlet{myYellow}{yellow}

\date{\today}

\begin{abstract}

 Symplectic gauge theories coupled to matter fields lead to symmetry enhancement 
 phenomena that have potential applications 
 in such diverse contexts as  composite Higgs,  top partial compositeness, 
 strongly interacting dark matter, and dilaton-Higgs models.
 These theories are also interesting on theoretical grounds, for example in reference to the
  approach to the large-$N$ limit.
A particularly compelling research   aim is
 the determination of the extent of the conformal 
 window in gauge theories with symplectic groups coupled to matter, 
for different groups and for field content consisting of fermions transforming in different 
 representations. Such determination would have 
 far-reaching implications,
 but requires
overcoming  huge  technical challenges.
 
Numerical studies based on lattice field theory can provide 
the quantitative information necessary to this endeavour. 
We  developed new software to
implement symplectic  groups in the Monte Carlo algorithms
within the Grid framework.
In this paper, we focus most of our attention on  the $Sp(4)$ lattice gauge theory coupled to
four (Wilson-Dirac) fermions transforming in the 2-index antisymmetric
representation, as a case study.
We discuss an extensive catalogue of technical tests of the algorithms and 
present  preliminary measurements to set the stage for future large-scale numerical investigations.
We also include the scan of parameter space of all asymptotically free $Sp(4)$ lattice gauge theories coupled to varying number of fermions transforming in the antisymmetric representation.

\end{abstract}
\preprint{PNUTP-23/A03}
\preprint{CTPU-PTC-23-26}
\maketitle

\tableofcontents

\section{Introduction}
\label{Sec:introduction}

Gauge theories with symplectic group, $Sp(2N)$, in four space-time dimensions have been proposed as the microscopic origin of several  new physics models that stand out in the literature for their simplicity and elegance. 
We  list some compelling examples later in this introduction.
Accordingly, lattice field theory methods have been deployed to obtain numerically 
a first quantitative characterisation of  the strongly coupled dynamics of such gauge theories~\cite{Bennett:2017kga,Lee:2018ztv,Bennett:2019jzz,Bennett:2019cxd,Bennett:2020hqd,Bennett:2020qtj,Lucini:2021xke,Bennett:2021mbw,Bennett:2022yfa,Bennett:2022gdz,Bennett:2022ftz,AS,Lee:2022elf,Hsiao:2022kxf,Bennett:2023rsl,Maas:2021gbf,Zierler:2021cfa,Kulkarni:2022bvh,Bennett:2023wjw}. Different regions of lattice parameter space have been explored; by 
varying the rank of the group, $N$, the number, $N_{\rm f, as}$, and  mass, $m^{\rm f, as}$, of (Dirac) fermions  transforming in the fundamental (${\rm f}$) and 2-index antisymmetric (${\rm as}$) representation, one can tabulate the properties of these theories. And, after taking infinite volume and continuum limits, the results can be used by model builders, phenomenologists, and field theorists working on  potential applications. 

A prominent role in the recent literature is played by the theory with $N=2$, $N_{\rm f}=2$, and
$N_{\rm as}=3$. It  gives rise, at low energies, to the effective field theory (EFT) entering   the minimal Composite Higgs model (CHM) that is amenable to lattice studies~\cite{Barnard:2013zea},\footnote{The literature on CHMs in which the Higgs fields emerge as 
pseudo-Nambu-Goldstone bosons (PNGBs) from the spontaneous breaking of the approximate global symmetries of a new, strongly coupled theory~\cite{Kaplan:1983fs,
Georgi:1984af,Dugan:1984hq}, is vast. See, e.g., the reviews in Refs.~\cite{
Panico:2015jxa,Witzel:2019jbe,
Cacciapaglia:2020kgq}, the summary tables in Refs.~\cite{
Ferretti:2013kya,Ferretti:2016upr,Cacciapaglia:2019bqz},
and the selection of papers in Refs.~\cite{
Katz:2005au,Barbieri:2007bh,
Lodone:2008yy,Gripaios:2009pe,Mrazek:2011iu,Marzocca:2012zn,Grojean:2013qca,Cacciapaglia:2014uja,
Ferretti:2014qta,Arbey:2015exa,Cacciapaglia:2015eqa,Feruglio:2016zvt,DeGrand:2016pgq,Fichet:2016xvs,
Galloway:2016fuo,Agugliaro:2016clv,Belyaev:2016ftv,Csaki:2017cep,Chala:2017sjk,Golterman:2017vdj,
Csaki:2017jby,Alanne:2017rrs,Alanne:2017ymh,Sannino:2017utc,Alanne:2018wtp,Bizot:2018tds,
Cai:2018tet,Agugliaro:2018vsu,Cacciapaglia:2018avr,Gertov:2019yqo,Ayyar:2019exp,
Cacciapaglia:2019ixa,BuarqueFranzosi:2019eee,Cacciapaglia:2019dsq,Cacciapaglia:2020vyf,
Dong:2020eqy,Cacciapaglia:2021uqh,Banerjee:2022izw,Ferretti:2022mpy}
and 
Refs.~\cite{
Contino:2003ve,Agashe:2004rs,Agashe:2005dk,Agashe:2006at,
Contino:2006qr,Falkowski:2008fz,Contino:2010rs,Contino:2011np,Caracciolo:2012je,Erdmenger:2020lvq,Erdmenger:2020flu,Elander:2020nyd,Elander:2021bmt,Elander:2021kxk,Elander:2023aow,Erdmenger:2023hkl}.}  and also realises top (partial)  compositeness~\cite{Kaplan:1991dc}
(see also Refs.~\cite{Grossman:1999ra,Gherghetta:2000qt}). It  hence provides an economical way of explaining the microscopic origin of the two heaviest particles in the standard model, the Higgs boson and the top quark,
singling them out as portals to new physics.

The $Sp(2N)$ gauge theories with $N_{\rm f}=2$ and $N_{\rm as}=0$ find  application also
in the simplest realisations of the strongly interacting massive particle (SIMP) scenario for dark matter~\cite{Hochberg:2014dra,Hochberg:2014kqa,Hochberg:2015vrg,Bernal:2017mqb,Berlin:2018tvf,
Bernal:2019uqr,Tsai:2020vpi,Kondo:2022lgg,Bernal:2015xba}. They can address observational puzzles such as the  {\it `core vs. cusp'}~\cite{deBlok:2009sp} and {\it `too big to fail'}~\cite{Boylan-Kolchin:2011qkt} problems. In addition, they might  have profound implications in the physics of the early universe and be  testable in present and  future gravitational wave experiments~\cite{Seto:2001qf,
 Kawamura:2006up,Crowder:2005nr,Corbin:2005ny,Harry:2006fi,
 Hild:2010id,Yagi:2011wg,Sathyaprakash:2012jk,Thrane:2013oya,
 Caprini:2015zlo,
 LISA:2017pwj,
 LIGOScientific:2016wof,Isoyama:2018rjb,Baker:2019nia,
 Brdar:2018num,Reitze:2019iox,Caprini:2019egz,Maggiore:2019uih}. This is because they can give rise to a relic stochastic background of gravitational 
waves~\cite{Witten:1984rs,Kamionkowski:1993fg,Allen:1996vm,Schwaller:2015tja, Croon:2018erz,Christensen:2018iqi},
that are the current subject of active study~\cite{Huang:2020crf,Halverson:2020xpg,Kang:2021epo}. 

On a more abstract, theoretical side, in $Sp(2N)$ Yang-Mills theories one can compute numerically the spectra of glueballs and strings~\cite{Lucini:2001ej,Lucini:2004my,Lucini:2010nv,Lucini:2012gg,
Athenodorou:2015nba,Lau:2017aom,Hernandez:2020tbc,Athenodorou:2021qvs,Yamanaka:2021xqh,
Bonanno:2022yjr},  as well as the topological charge and 
susceptibility~\cite{Luscher:1981zq,Campostrini:1989dh,DelDebbio:2002xa,Lucini:2004yh,DelDebbio:2004ns,Luscher:2010ik,Panagopoulos:2011rb,Bonati:2015sqt,Bonati:2016tvi, Ce:2016awn,Alexandrou:2017hqw,Bonanno:2020hht,Borsanyi:2021gqg,Cossu:2021bgn,
Teper:2022mmj,Bonanno:2022vot,Bonanno:2022hmz}.
This allows for a comparison with other gauge groups ($SU(N_c)$ in particular),
by means of which to test non-perturbative ideas about field theories and their approach to the large-$N_c$ limit---see,
e.g., Refs.~\cite{Bochicchio:2016toi,Bochicchio:2013sra,
Hong:2017suj,Bennett:2020hqd,Bennett:2022gdz}.
Indeed, even the pioneering lattice study of symplectic theories in Ref.~\cite{Holland:2003kg}
was performed to the purpose of better characterising on general grounds the deconfinement phase transition.

A special open problem is that of the highly non-trivial determination of the extent of the conformal window in strongly coupled gauge theories with matter field content.
It has both theoretical and phenomenological implications, of general interest to model-builders, phenomenologists, and field theorists alike.
Particular attention has been so far paid to $SU(N_c)$ theories, more than 
$Sp(2N)$ (with $N>1$) ones.
 Let us pause and explain what the problem is, on general grounds. Robust perturbation-theory arguments show that if the number of matter fields is large enough---but not so much as to spoil asymptotic freedom---gauge theories can be realised in a conformal phase. This is the case when long distance physics is governed by a fixed point of the renormalisation group (RG) evolution~\cite{Caswell:1974gg,Banks:1981nn}, and the fixed point is described by a conformal field theory (CFT). It is reasonable to believe that such fixed points may exist also outside the regime of validity of perturbation theory, 
when the number of matter fields is smaller.
What is the smallest number of fermions for which the theory still admits a fixed point, rather than confining in the infrared (IR), is an open question.  
While gaining some control over non-perturbative physics is possible in supersymmetric theories (see Ref.~\cite{Intriligator:1995au} and references therein),  the non-supersymmetric ones are the subject of a rich and fascinating literature~\cite{Appelquist:1988yc,Cohen:1988sq,Sannino:2004qp,Dietrich:2006cm,Ryttov:2007cx,Pica:2010mt,Pica:2010xq,Kim:2020yvr,Lee:2020ihn}, part of which uses perturbative instruments and high-loop expansions~\cite{Baikov:2016tgj,Herzog:2017ohr,Ryttov:2016ner,Ryttov:2016hdp,
Ryttov:2016asb,Ryttov:2016hal,Ryttov:2017toz,
Ryttov:2017kmx,Ryttov:2017dhd,Gracey:2018oym,Ryttov:2018uue,Ryttov:2020scx,Ryttov:2023uzc},
but there is no firm agreement on the results---we include a brief overview of work in this direction, in the body of the paper.

Knowledge of the extent of the conformal window also has relevant phenomenological  implications. Various arguments 
suggest that at the lower edge of the conformal window, the anomalous dimensions of the CFT operators might be so large as to invalidate naive dimensional analysis (NDA) expectations for the scaling of observable quantities~\cite{Cohen:1988sq,Kaplan:2009kr}. And it has been speculated that this might affect even  confining theories that 
live outside the conformal window, with applications to technicolor, 
CHMs, top (partial)  compositeness, SIMP dark matter (e.g., see 
Refs.~\cite{Panico:2015jxa,Witzel:2019jbe,Cacciapaglia:2020kgq,Chivukula:2000mb,Lane:2002wv,
Hill:2002ap,Martin:2008cd,Sannino:2009za,Piai:2010ma} and references therein).

Lattice studies of the extent of the conformal window
 have mostly focused on $SU(N_c)$ groups, with fermion matter in various representations
of the gauge group.\footnote{See for instance the review in Ref.~\cite{Rummukainen:2022ekh}, and references therein, in particular Refs.~\cite{Catterall:2007yx,DelDebbio:2008zf,Hietanen:2008mr,Appelquist:2009ty,Hietanen:2009az,
DelDebbio:2009fd,LSD:2009yru,Bursa:2009we,Fodor:2009ff,DelDebbio:2010hx,DeGrand:2010na,
Patella:2010dj,Hayakawa:2010yn,Fodor:2011tu,Bursa:2011ru,Appelquist:2011dp,DeGrand:2011cu,
Karavirta:2011zg,DeGrand:2012qa,Appelquist:2012nz,Lin:2012iw,Aoki:2012eq,Cheng:2013eu,
DeGrand:2013uha,Hasenfratz:2013eka,LSD:2014nmn,Lombardo:2014pda,Hasenfratz:2014rna,Athenodorou:2014eua,
Fodor:2015baa,
Fodor:2015zna,Rantaharju:2015yva,Rantaharju:2015cne,Fodor:2016zil,Arthur:2016dir,Arthur:2016ozw,Athenodorou:2016ndx,Hasenfratz:2016dou,
Leino:2017lpc,Leino:2017hgm,Amato:2018nvj,Fodor:2018tdg,Hasenfratz:2019dpr,
Hasenfratz:2020ess,LatticeStrongDynamics:2020uwo,Lopez:2020van,
Athenodorou:2021wom,Bennett:2021ivn,Bergner:2022hoo,Bennett:2022bhc,Bergner:2022trm,Hasenfratz:2023wbr}.} 
Closely related to these studies is the emergence, in $SU(3)$ gauge theories with eight (Dirac) 
fermions transforming in the fundamental representation~\cite{LatKMI:2014xoh,Appelquist:2016viq,LatKMI:2016xxi,Gasbarro:2017fmi,LatticeStrongDynamics:2018hun,LatticeStrongDynamicsLSD:2021gmp,Hasenfratz:2022qan,LSD:2023uzj,LatticeStrongDynamics:2023bqp}, or (Dirac) fermions transforming in the 2-index symmetric representation~\cite{Fodor:2012ty,
	Fodor:2015vwa,
	Fodor:2016pls,Fodor:2017nlp,Fodor:2019vmw,Fodor:2020niv},  of numerical evidence
 pointing to the existence of a light  isosinglet scalar state, that is tempting to identify with the dilaton,
the PNGB associated with dilatations.

It has been predicted long ago that a light dilaton should exist in strongly coupled,
confining theories living  in proximity of the lower end of the conformal window~\cite{Leung:1985sn,Bardeen:1985sm,Yamawaki:1985zg}, and the EFT description of such state 
has a remote historical origin~\cite{Migdal:1982jp,Coleman:1985rnk}. It might have huge consequences in extensions of the standard model~\cite{Goldberger:2008zz}. A plethora of phenomenological studies exists on the dilaton (see, for example, Refs.~\cite{Hong:2004td,Dietrich:2005jn,Hashimoto:2010nw,Appelquist:2010gy,Vecchi:2010gj,Chacko:2012sy,Bellazzini:2012vz,Bellazzini:2013fga,Abe:2012eu,Eichten:2012qb,Hernandez-Leon:2017kea,CruzRojas:2023jhw} and references therein).
The $SU(3)$ lattice evidence for the existence of this state has triggered renewed interest in the dilaton effective field theory (dEFT),  which combines the chiral Lagrangian description of the PNGBs associated with the internal global symmetries of the system, with the additional, light scalar, interpreted as a dilaton~\cite{Matsuzaki:2013eva,Golterman:2016lsd,Kasai:2016ifi,Hansen:2016fri,Golterman:2016cdd,Appelquist:2017wcg,Appelquist:2017vyy,Golterman:2018mfm,Cata:2019edh,Cata:2018wzl,Appelquist:2019lgk,Golterman:2020tdq,Golterman:2020utm,Appelquist:2020bqj,Appelquist:2022qgl,Appelquist:2022mjb}.

The aforementioned lattice studies of symplectic theories, motivated by  CHMs and SIMPs, can be carried out with comparatively modest resources, and using lattices of modest sizes, because they require exploring the intermediate mass range  for the mesons in the theory. By contrast, the study of the deep-IR properties of $Sp(2N)$ gauge theories  requires investigating the low mass regime of the fermions, for which one needs lattices and ensembles big enough to overcome potentially large  finite size effects and long autocorrelation times. The supercomputing demands (both on hardware and software) of these calculations
are such that a new dedicated set of instruments, and a long-term research strategy, is needed to make these investigations feasible.
With this paper, we make the first, propaedeutic, technical steps on the path towards determining on the lattice the extent of the conformal window in theories with $Sp(2N)$ group, for $N>1$. 

To this end, we elected to build, test, and make publicly available new software \cite{githubsp2n}, that supplements previous releases of the Grid library~\cite{Boyle:2015tjk,Boyle:2016lbp,Yamaguchi:2022feu,github},  by adding to it new
  functionality specifically designed to handle $Sp(2N)$ theories with matter fields in multiple representations. 
  The resulting software takes advantage of all the features offered by the modularity and flexibility of Grid, in particular its ability to work both on CPU- as well as GPU-based architectures.    We present two types of preliminary results relevant to this broader endeavour:  technical tests of the algorithm and of the physics outcomes   are supplemented by preliminary analyses, conducted on coarse lattices, of the parameter space of the lattice theory. The latter set the
stage for future large-scale numerical studies, by identifying the regions of parameter space connected to continuum physics. The former are 
intended to validate the software, and test its performance for symplectic theories on machines with GPU architecture.
Unless otherwise specified, we use the $Sp(4)$ theory, coupled to $N_{\rm as}=4$ Wilson-Dirac fermions transforming in the 2-index antisymmetric representation, as a case study. The lessons we learn from the results we report have general validity and applicability.

 This paper is organised as follows. We start by defining the $Sp(2N)$ gauge theories of interest in Sect.~\ref{Sec:gauge},
 both in the continuum and on the lattice. We also summarise briefly the current understanding of
the extent of the conformal window in these theories.
Section~\ref{Sec:grid} discusses the software implementation of $Sp(2N)$ on Grid,
and the basic tests we performed on the algorithm. 
In Sect.~\ref{Sec:2} we concentrate on lattice theories in which the fermions do not contribute to the dynamics, focusing both on the Yang-Mills theory and the quenched approximation.
New results about the bulk structure of all the $Sp(4)$ theories coupled to (Wilson-Dirac) fermions
 transforming in the 2-index antisymmetric representation can be found in Sect.~\ref{Sec:2-4}, while Sect.~\ref{Sec:sst} discusses scale setting (Wilson flow) and topology.
A brief summary and outlook concludes the paper, in Sect.~\ref{Sec:conclusions}.
Additional technical details are relegated to the appendix.
\section{Gauge theories with symplectic group}
\label{Sec:gauge}

The $Sp(2N)$ continuum field theories of interest (with $N>1$), written in Minkowski space with signature mostly `$-$', have the following Lagrangian density (we borrow notation and conventions from Ref.~\cite{Bennett:2019cxd}):
\beqs
{\cal L}&=& -\frac{1}{2} \Tr \ G_{\mu\nu} G^{\mu\nu}
\,+\,\frac{1}{2}\sum_i^{N_{\rm f}}\left(i\overline{Q^{i}}_a \gamma^{\mu}\left(D_{\mu} Q^i\right)^a
\,-\,i\overline{D_{\mu}Q^{i}}_a \gamma^{\mu}Q^{i\,a}\right)\,-\,m^{\rm f}\sum_i^{N_{\rm f}}\overline{Q^i}_a Q^{i\,a}+\nonumber\\
&&
\,+\,\frac{1}{2}\sum_k^{N_{\rm as}}\left(i\overline{\Psi^{k}}_{ab} \gamma^{\mu}\left(D_{\mu} \Psi^k\right)^{ab}
\,-\,i\overline{D_{\mu}\Psi^{k}}_{ab} \gamma^{\mu}\Psi^{k\,ab}\right)\,-\,m^{\rm as} \sum_k^{N_{\rm as}}\overline{\Psi^k}_{ab} \Psi^{k\,ab}\,.
\label{eq:lagrangian}
\eeqs
The fields $Q^{i\,a}$, with $i=1,\,\cdots,\,N_{\rm f}$, are Dirac fermions that transform in the fundamental representation of $Sp(2N)$, 
as indicated by the index $a=1,\,\cdots,\,2N$, while the 
$\Psi^{k\,ab}$ ones, with $k=1,\,\cdots,\,N_{\rm as}$, transform in the 2-index antisymmetric 
representation of the gauge group.
The covariant derivatives are defined by making use of the transformation 
properties under the action of an element $U$ of the 
$Sp(2N)$ gauge group, according to which
\beqs
Q\rightarrow U Q\,,~~~ {\rm and} ~~~ \Psi \rightarrow U \Psi U^{\mathrm{T}}\,.
\eeqs
They can be written in terms of the gauge field $A_{\mu}\equiv A_{\mu}^a t^a$,
where $t^a$ are the generators of $Sp(2N)$, normalised so that $\Tr\ t^at^b=\frac{1}{2}\delta^{ab}$,
to read as follows:
\beqs
D_{\mu} Q^i&=& \partial_{\mu} Q^i \,+\,i g A_{\mu} Q^{i}\,,\\ 
D_{\mu} \Psi^j&=& \partial_{\mu} \Psi^j \,+\,i g A_{\mu} \Psi^{j}\,+\,i g \Psi^{j} A_{\mu}^{\mathrm{T}}\,,
\eeqs
where $g$ is the gauge coupling. The field-strength tensor is given by
\beqs
G_{\mu\nu}&\equiv& \partial_{\mu}A_{\nu}-\partial_{\nu}A_{\mu} + i g \left[A_{\mu}\,,\,A_{\nu}\right]\,,
\eeqs
where $[\cdot \,, \cdot ]$ is the commutator.

The form of Eq.~(\ref{eq:lagrangian})   makes it easy to show that 
the $SU(N_{\rm f})_L\times SU(N_{\rm f})_R$ and 
$SU(N_{\rm as})_L\times SU(N_{\rm as})_R$ 
global symmetries acting on the flavor indexes of $Q^i$ and $\Psi^k$, respectively, are enhanced to $SU(2N_{\rm f})$ and $SU(2N_{\rm as})$. By rewriting  Eq.~(\ref{eq:lagrangian})  in
 terms of 2-component fermions~\cite{Lewis:2011zb,Bennett:2019cxd} , 
 \\
\begin{eqnarray}
    Q^{i\,a} = \begin{pmatrix} q^{i\,a} \\ \Omega^{ab} \left( -\tilde{C}q^{i+2, *} \right)_b \end{pmatrix} \,, \;  \Psi^{k\, ab} = \begin{pmatrix} \psi^{k\, ab} \\ \Omega^{ab} \Omega^{bd} \left( -\tilde{C}q^{k+3, *} \right)_{cd} \end{pmatrix}\,,
\end{eqnarray}
($\tilde{C} = -i \tau^2$, $\tau^2$ is the second Pauli matrix) we get the Lagrangian where the global symmetries are manifest\\
\beqs
{\cal L}&=& -\frac{1}{2} \Tr \ G_{\mu\nu} G^{\mu\nu}
\,+\,\frac{1}{2} \sum_{j}^{2N_{\rm f}} \left( i\left( q^{j} \right)^{\dagger}_a \bar{\sigma}^{\mu} \left( D_\mu q^{j} \right)^a - i\left( D_\mu q^{j} \right)^{\dagger}_a \bar{\sigma}^{\mu} \left( q^{j} \right)^a \right) +\nonumber\\  &&
\,-\, \dfrac{1}{2} m^{\rm f} \sum_{j,\,k}^{2N_{\rm f}} \Omega_{jk} \left( q^{j\,a\,T} \Omega_{ab} \tilde{C} q^{k\,b} - \left(q^{j}\right)^{\dagger}_{a} \Omega^{ab} \tilde{C} \left(q^{k \, *} \right)_b \right)  +\nonumber\\  &&
\,+\, \dfrac{1}{2} \sum_{k}^{2N_{\rm as}} \left( i \left( \psi^k\right)^\dagger_{ab} \bar{\sigma}^\mu \left( D_\mu \psi^k \right)^{ab} - i \left( D_\mu \psi^k\right)^\dagger_{ab} \bar{\sigma}^\mu \left( \psi^k \right)^{ab} \right)   +\nonumber\\  &&
\,-\ \dfrac{1}{2} m^{\rm as} \sum_{j,\,k}^{2N_{\rm as}} \omega_{jk} \left( \psi^{j \, ab \, T} \Omega_{ac} \Omega_{bd} \tilde{C} \psi^{k \, cd} - \left( \psi^{j} \right)^{\dagger}_{ab} \Omega^{ac} \Omega^{bd} \tilde{C} \left( \psi^{k \, *} \right)_{cd} \right) \,,
\label{eq:lagrangian_twocomponent}
\eeqs\\
where we defined $\bar{\sigma}^\mu \equiv \left( \mathbb{1}_{2 \times 2} , \tau^i \right)$ and $\omega_{jk} = \omega^{jk} \equiv \begin{pmatrix}
\begin{matrix} \mathbb{0}\end{matrix} & \mathbb{1}_{N_{\rm as}} \\ \mathbb{1}_{N_{\rm as}} & \begin{matrix} \mathbb{0} \end{matrix}
\end{pmatrix}$. The antisymmetric matrix $\Omega$ has the same form, as defined in Eq.~(\ref{eq:Omega_matrix}) of Appendix A, for both the gauge and fundamental flavour symmetries, but the indices run with $a=1,\cdots,2N$ for the former and with $j=1,\cdots,2N_{\rm f}$ for the latter.
 
The mass terms break the symmetries to the maximal $Sp(2N_{\rm f})$ and $SO(2N_{\rm as})$ subgroups.
Bilinear fermion condensates  arise non-perturbatively,
breaking the symmetries according to the same pattern, and hence one expects the presence of 
$N_{\rm f}(2N_{\rm f} -1)-1$ PNGBs in the $(\rm f)$ sector (for $N_f>1$), and $N_{\rm as}(2N_{\rm as} +1)-1$  in the $(\rm as)$ sector.

The main parameters governing the system are hence $N$, $N_{\rm f}$, and $N_{\rm as}$, and in most of the paper we 
refer to the theory with $N=2$, $N_{\rm f}=0$, and $N_{\rm as}=4$ as a case study.
The running coupling, $g$,
obeys a renormalisation group equation (RGE)
in which the beta function at the 1-loop order is  scheme-independent, 
\beqs
\beta&=&-\frac{g^3}{(4\pi)^2} b_1,
\eeqs
and is governed by the coefficient $b_1$, which for a non-Abelian theory coupled to Dirac fermions
can be written as
\beq
b_1
= \frac{11}{3}C_2(G)-\frac{4}{3}N_{\rm f} \frac{d_{\rm f}}{d_G}C_2({\rm f})
-\frac{4}{3}N_{\rm as}  \frac{d_{\rm as}}{d_G}C_2({\rm as})\,\\
\eeq
and, specifically for $Sp(2N)$ groups, becomes 
\beq
b_1=\frac{11}{3}(N+1)
-\frac{2}{3}N_{\rm f}
-\frac{4}{3}N_{\rm as} \frac{N(2N-1)-1}{N(2N+1)}
N\,.
\eeq
The coefficients $C_2 (G)$, $C_2 (\rm{f})$, $C_2 (\rm{as})$ are quadratic Casimir operators in the adjoint, fundamental and antisymmetric representations, while $d_G$, $d_{\rm f}$, $d_{\rm as}$ are the dimensions of these representations, respectively. We  restrict attention to  asymptotically free theories, for which $b_1$ is  positive. For $Sp(2N)$ theories with $N_{\rm f}=0$, this  requirement 
sets the upper bound $N_{\rm as}<\frac{11(N+1)}{4(N-1)}$,
which for $N=2$ yields $N_{\rm as}<33/4$---perturbatively, ${\rm as}$-type
fermions make double the contribution of ${\rm f}$-type ones, in $Sp(4)$.
The spectrum of mesons depends  on the mass, $m^{\rm f,as}$, of the  fermions, 
by varying which we can test which of the following three possible classes
 the theory falls into.

\begin{enumerate}[label=(\roman*)]
\item The theory confines, similarly to Yang-Mills theories. One expects to find a gapped spectrum, 
and a set of  PNGBs that become parametrically light in respect to other states, when $m^{\rm f,as}\rightarrow 0$. The small mass and momentum regime is described by chiral perturbation theory ($\chi$PT)~\cite{Coleman:1969sm,Weinberg:1978kz,Gasser:1984gg,Leutwyler:1993iq}.
\item The theory is IR conformal. In this case, a gap arises only because of the presence of the mass terms,
and would disappear into a continuum for  $m^{\rm f,as}\rightarrow 0$. The spectrum and spectral density exhibit scaling,
in the form described for example in Refs.~\cite{Luty:2008vs,DelDebbio:2009fd,DeGrand:2009hu,DelDebbio:2010hu,DelDebbio:2010ze,DelDebbio:2010jy,Patella:2012da}---see also Ref.~\cite{Hasenfratz:2023sqa}.
\item The theory is confining, but has near-conformal dynamics. As in the confining case, when 
$m^{\rm f,as}\rightarrow 0$ one finds massless PNGBs. An additional  isosinglet scalar state, the dilaton, 
is also light, compared to  the other mesons, and long distance physics is described by  dEFT~\cite{Matsuzaki:2013eva,Golterman:2016lsd,Kasai:2016ifi,Hansen:2016fri,Golterman:2016cdd,Appelquist:2017wcg,Appelquist:2017vyy,Golterman:2018mfm,Cata:2019edh,Cata:2018wzl,Appelquist:2019lgk,Golterman:2020tdq,Golterman:2020utm,Appelquist:2020bqj,Appelquist:2022qgl,Appelquist:2022mjb}---see also the discussions in Refs.~\cite{Crewther:2020tgd,DelDebbio:2021xwu,Zwicky:2023bzk}, and references therein.
\end{enumerate}

\subsection{The conformal window}
\label{Sec:cw}

The three possible classes of gauge theories described above are determined by whether the theory  is, respectively, far outside, inside or just outside the boundary of the conformal window.
The determination of the conformal window is tantamount to showing the existence of the IR fixed point at non-zero coupling 
so that the theory is interacting and IR conformal. We provide here some more detail and information about this
challenging endeavour and what is known to date, starting from perturbative arguments.
The coefficient of the (scheme-independent) $2$-loop RG beta function, $b_2$, which is found to be, for generic non-abelian gauge theories,
\beq
b_2=
\frac{34}{3}C_2(G)^2-\frac43\left(5C_2(G)+3C_2(\rm {f})\right) \frac{d_{\rm f}}{d_G} C_2(\rm{f}) N_{\rm {f}} -\frac43 \left(5C_2(G)+3C_2(\rm {as})\right) \frac{d_{\rm as}}{d_G} C_2(\rm{as}) N_{\rm {as}}\,, \\	
\eeq
and for $Sp(2N)$ groups reduces to
\beq
b_2=\frac{34}{3}(N+1)^2
-\frac{2}{3} N_{\rm f} \left[  5(N+1) + \frac34 \left( 2N + 1 \right) \right]
-\frac{4}{3}N_{\rm as} \left[ 3N + 5\left( N + 1\right) \right] \frac{N(2N-1)-1}{2N+1}
\,,
\eeq
When $b_2$ is negative,  one finds that  for a positive and sufficiently small value of $b_1$,   
a perturbative IR fixed point at coupling $\alpha_{\rm IR} \simeq \alpha_{\rm BZ} = -4\pi b_1/b_2 \ll 1$ 
arises. This is referred to as a Banks-Zaks (BZ) fixed point~\cite{Caswell:1974gg,Banks:1981nn}.
The upper bound of the conformal window therefore coincides with that of asymptotically free theories, given by $b_1=0$. 

The determination of the lower bound of the conformal window is hindered by the vicinity of the strong coupling regime. 
To see this, one can fix the value of $N$ and decrease the number of flavors $N_{\rm{f, as}}$. The coefficient $b_2$ then becomes less negative and eventually approaches zero, while $b_1$ remains finite and positive. 
Accordingly, the coupling at the (perturbative) BZ fixed point, $\alpha_{\rm BZ}$, becomes larger and larger and the perturbative analysis of the $\beta$ function is no longer reliable.  
Despite such inherent limitations, several (approximate) analytical methods have been proposed to estimate the critical 
value $N_{\rm{f, as}}^{\rm cr}$ 
corresponding to the lower edge of the conformal window.
We now briefly summarise known results, for the theories of interests, that can be used to guide 
dedicated studies using non-perturbative numerical techniques, such as those based on lattice field theory.

Let us start by setting $N_{\rm f} = 0$ and varying $N_{\rm{as}}$. A na\'ive estimate can be derived by taking the perturbative 2-loop beta function to hold beyond perturbation theory, 
using it to compute $N_{\rm as}^{\rm BZ,\,cr}$, and
assuming that the fixed point disappears when $\alpha_{\rm BZ}\rightarrow \infty$, or equivalently by looking for solutions of the condition $b_2\rightarrow 0$. Doing so yields $N_{\rm as}^{{\rm BZ},\,{\rm cr}}\simeq 3.7$ for $Sp(4)$. 
This approach can  be systematically improved by including
 higher-order loops, up to ${\ell}_{\rm max}>2$, in the expansion of the beta function $\beta(\alpha)$. One
then seeks values of $N_{\rm as}$ for which $\alpha_{\rm IR}\rightarrow \infty$, with $\alpha_{\rm IR}$ determined 
by solving $\beta(\alpha)\equiv -2\alpha \sum_{\ell =1}^{\ell_{\rm max}}b_\ell \left(\frac{\alpha}{4\pi}\right)^\ell=0$. 
In particular, one finds $N_{\rm as}^{\rm 4-loop,\, cr} \simeq 4.1$ from the perturbative beta function at four loops in the $\overline{\rm MS}$-scheme \cite{vanRitbergen:1997va}. 
It should be noted, however, that the results are affected by uncontrolled systematics,
 since the coefficients, $b_{\ell}$, of the beta function, $\beta(\alpha)$, depend
  on the renormalisation scheme at three or higher loops, when $\ell \geq 3$. 

An alternative approach makes use of the Schwinger-Dyson (SD) equation in the ladder approximation,  
in which case conformality is assumed to be lost when $\alpha_{\rm IR}\equiv \alpha^{\rm cr}$, with $\alpha^{\rm cr}=\pi/3C_2(R)$, 
which yields $N_{\rm as}^{\rm SD}\simeq 6$ for $Sp(4)$.  
Going beyond the perturbative coupling expansion, 
a conjectured all-orders beta function $\beta^{\rm all-orders}(\alpha)$ \cite{Ryttov:2007cx}, which involves the first two universal coefficients of $\beta(\alpha)$ and the anomalous dimension of fermion bilinear operator,  $\gamma_{\bar{\psi}\psi}(\alpha)$, 
has been proposed.\footnote{
A modified version of the all-orders beta function can also be found in Ref.~\cite{Pica:2010mt}. 
} In this case,  the conformal window is determined by solving the condition
$\beta^{\textnormal{all-orders}}=0$ with the physical input for the value of $\gamma_{\bar{\psi}{\psi}}$ at the IR fixed point. 
For  $\gamma_{\bar{\psi}\psi}=1$, one finds $N_{\rm as}^{\rm all-orders,\,cr}\simeq 5.5$ for $Sp(4)$.\footnote{
This choice for $\gamma_{\bar{\psi}\psi}$ has been argued to be the critical condition associated with the chiral phase transition 
through the IR and UV fixed point merger \cite{Kaplan:2009kr}, and by matching smoothly to the chiral phase with pions~\cite{Zwicky:2023bzk}. A less common  choice is to set  $\gamma_{\bar{\psi}\psi}=2$, as suggested by unitarity considerations~\cite{Mack:1975je}.
}

       \begin{figure}
       
          \begin{center}
        \includegraphics[width=0.5\linewidth]{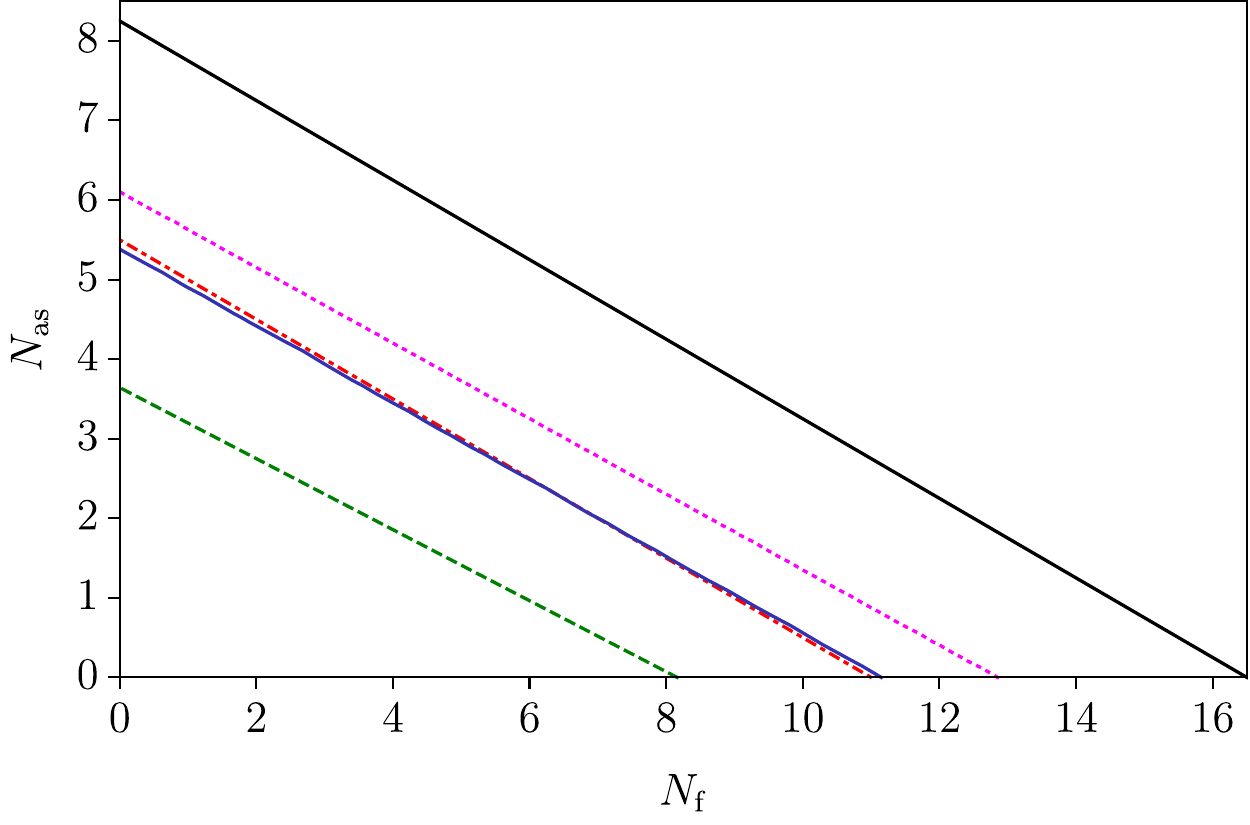}
      \caption{Estimates of the extent of the conformal window in $Sp(4)$ theories coupled to $N_{\rm f}$ Dirac fermions transforming in the  fundamental and $N_{\rm as}$ in the 2-index  antisymmetric representation. 
      The black solid line denotes the upper bound of the conformal window, while different colored and shaped lines denote alternative analytical estimates 
      of the lower bound, obtained with different approximations. 
      The dashed line is obtained by imposing the constraint $b_2(N_{\rm f},\,N_{\rm as})=0$. 
The dot-dashed line is the result of the all-order beta function with the assumption that the anomalous dimensions of the fermion bilinears are
$\gamma_{\bar{\Psi}\Psi}=\gamma_{\bar{Q}Q}=1$. 
The dotted line is the result of the SD analysis. The BZ expansion
leads to the lower (blue)
solid line. Details about these approximations can be found in the main text and in the reference list. 
       }
           \label{Fig:cw_sp4}
          \end{center}
       \end{figure}

More recently, the scheme-independent BZ expansion in the small parameter $\Delta_{N_{\rm{as}}}=N_{\rm{as}}^{\rm AF}-N_{\rm{as}}^{\rm IR}$
has been extensively  applied to the determination of physical quantities such as the anomalous dimension, 
$\gamma_{\bar{\psi}\psi}$, 
at the IR fixed point---see Ref.~\cite{Ryttov:2017kmx} and refs. therein.  
In Ref.~\cite{Kim:2020yvr}, the authors determined the lower edge of the conformal window 
by imposing the critical condition of $\gamma_{\bar{\psi}\psi} (2-\gamma_{\bar{\psi}\psi})=1$. 
This condition is identical to $\gamma_{\bar{\psi}\psi}=1$ at infinite order, 
but displays better convergence at finite order in the $\Delta_{N_{\rm{as}}}$ expansion. 
The 4th order calculation yields $N_{\rm as}^{\gamma cc,\, {\rm cr}}\simeq 5.5$ for $Sp(4)$~\cite{Lee:2020ihn}.

These analytical approaches can be extended to determine the conformal window for the theory containing fermions in the multiple representations, $\{R_1,\,R_2,\,\cdots,\,R_k\}$, 
in which case the upper and lower bounds of the conformal window are described by $(k-1)$-dimensional hyper-surfaces. 
For the $Sp(4)$ theories of interest  with $N_{\rm f}$ Dirac fermions transforming in the 
fundamental and $N_{\rm as}$ in the 2-index  antisymmetric representation, 
the results are summarised in Fig.~\ref{Fig:cw_sp4}.\footnote{
The figure is basically the same as the analogous one found in Ref.~\cite{Kim:2020yvr}, except that 
the input for the all-orders beta function analysis has been changed to $\gamma_{\bar{\Psi}\Psi}=\gamma_{\bar{Q}Q}=1$.  
The parameter space has also been extended and the notation adapted to the conventions of this paper. 
} 
The upper bound is determined by the condition $b_1(N_{\rm f},\,N_{\rm as})=0$. 
The various alternative determinations of the lower bound are estimated as follows.
The dashed line is obtained by setting $b_2(N_{\rm f},\,N_{\rm as})=0$. 
The dot-dashed line corresponds to the result of the all-order beta function with the input of $\gamma_{\bar{\Psi}\Psi}=\gamma_{\bar{Q}Q}=1$. 
The dotted and solid lines are the results of the SD analysis and the BZ expansion of $\gamma_{\bar{\Psi}\Psi}$ at the 3rd order in
 $\Delta_{N_{\rm f}(n_{\rm as})}$~\cite{Ryttov:2018uue} 
with the critical conditions applied to the antisymmetric fermions, $\alpha_{\rm BZ}=\alpha^{\rm cr}_{\rm as}=\pi/3 C_2({\rm AS})$ 
and $\gamma_{\bar{\Psi}\Psi} (2-\gamma_{\bar{\Psi}\Psi})=1$, respectively, 
as fermions in the higher representation are expected to condense first, resulting in the larger values of $\alpha^{\rm cr}$ and $\gamma_{\rm IR}$ \cite{Ryttov:2010hs}. 
It might be possible to make use of the five-loop computations in Refs.~\cite{Herzog:2017ohr, Baikov:2016tgj}, to further improve these estimations of the conformal window, but this goes
 beyond the purposes of this discussion.
For the purpose of  phenomenological applications, the most interesting physical quantities 
one would like to determine within the conformal window are 
the anomalous dimensions of fermion bilinear operators (mesons) and chimera baryon operators. 
Perturbative calculations of the former are available in the literature, up to  the 4th order of the coupling 
expansion~\cite{Chetyrkin:2016ruf} and at the 3rd order of the BZ expansion~\cite{Ryttov:2018uue}, 
while that of the latter is only available at the leading order in $\alpha$ \cite{BuarqueFranzosi:2019eee}.
All of these considerations, summarised in Fig.~\ref{Fig:cw_sp4}, offer some intuitive guidance for what can be expected, but non-perturbative instruments are needed to test these predictions and put Fig~\ref{Fig:cw_sp4} on firmer grounds.

\subsection{The lattice theory}
\label{Sec:lattice}

In presenting the lattice theory, we borrow again notation and conventions from Ref.~\cite{Bennett:2022yfa}. 
The theory is defined on a Euclidean, hypercubic, four-dimensional lattice with spacing $a$,
with $L/a$ sites in the space directions and $T/a$ in the time direction. 
The generic lattice site is denoted as $x$, and the link in
direction $\mu$ as $(x,\,\mu)$.
The
total number of sites is thus $\tilde V /a^4 =  T\times L^3/a^4$. Unless stated otherwise, in the following we set $L=T$. 
The action is the sum of two terms
\beq
S \equiv S_g + S_f\,,
\label{eq:lattice_action}
\eeq
where $S_g$ and $S_f$ are the gauge and fermion action, respectively.
Among the several choices for the former--the Iwasaki, Symanzik, DBW2 and Wilson gauge actions--for simplicity, we show our results using the Wilson action, defined as
\beq
S_g\equiv\beta \sum_x \sum_{\mu<\nu} \left(1-\frac{1}{2N} {\rm Re}\, {\Tr}\, 
{\cal P}_{\mu\nu}(x)
\right),
\label{eq:gauge_action}
\eeq
where ${\cal P}_{\mu\nu}(x)\equiv U_\mu (x) U_\nu (x+\hat{\mu}) U_\mu^\dagger(x+\hat{\nu}) U_\nu^\dagger(x)$ is
known as the \emph{elementary plaquette} operator, $U_{\mu}(x) \in Sp(2N)$ is the
\emph{link variable} defined on link $(x,\mu)$, and
$\beta\equiv 4N/g_0^2$, where $g_0$ is the bare gauge coupling.
For the fermions, we adopt the massive Wilson-Dirac action,
\beqs
S_f &\equiv& a^4 \sum_{j=1}^{N_{\rm f}}\sum_x \overline{Q}^j(x) D^{\mathrm{(f)}}_m Q^j(x)+
a^4 \sum_{j=1}^{N_{\rm as}}\sum_x \overline{\Psi}^j(x) D^{\mathrm{(as)}}_m \Psi^j(x)\,,
\label{eq:fermion_action}
\eeqs
where $Q^j$ and $\Psi^j$ are the fermion fields transforming, respectively, in
the fundamental and 2-index antisymmetric representation and $j$ is a flavor
index, while color and spinor indices are omitted for simplicity. 
The massive Wilson-Dirac operators in Eq.~(\ref{eq:fermion_action}) are defined as
\beqs
D^{\mathrm{(f)}}_m Q^j(x) &\equiv& (4/a+m^{\mathrm{f}}_0) Q^j(x)\label{eq:DiracF} \\
&&-\frac{1}{2a}\sum_\mu \nonumber
\left\{(1-\gamma_\mu)U^{\mathrm{(f)}}_\mu(x)Q^j(x+\hat{\mu})
+(1+\gamma_\mu)U^{\mathrm{(f)},\,\dagger}_\mu(x-\hat{\mu})Q^j(x-\hat{\mu})\frac{}{}\right\}\,,
\eeqs
and
\beqs
D^{\mathrm{(as)}}_m \Psi^j(x) &\equiv& (4/a+m^{\mathrm{as}}_0) \Psi^j(x) \label{eq:DiracAS}\\
&&-\frac{1}{2a}\sum_\mu \nonumber
\left\{(1-\gamma_\mu)U^{\mathrm{(as)}}_\mu(x)\Psi^j(x+\hat{\mu})
+(1+\gamma_\mu)U^{\mathrm{(as)},\,\dagger}_\mu(x-\hat{\mu})\Psi^j(x-\hat{\mu})\frac{}{}\right\}\,,
\eeqs
where $m_0^{\mathrm{f}}$ and $m_0^{\mathrm{as}}$ are the bare fermion masses in the
fundamental and $2$-index antisymmetric representation, and $U^{\mathrm{(f)}}_\mu(x)=U_\mu(x)$.
The link variables $U^{\mathrm{(as)}}_\mu(x)$ are defined as in Ref.~\cite{Bennett:2022yfa}, as follows:
\beq\label{eq:upgrade_Uf_to_Uas}
    U^{\mathrm{(as)}}_{\mu,\,(ab)(cd)} = 
    \Tr \left( e^{(ab)\,T} 
    U^{(\rm f)}_\mu e^{(cd)} 
    U^{(\rm f)\,T}_\mu  \right) \,,
\eeq
where $e^{(ab)}$ are the elements of an orthonormal basis in the $(N(2N-1)-1)$-dimensional
space of $2N\times2N$ antisymmetric and $\Omega$-traceless matrices, and
the multi-indices $(ab)$ run over the values 
$1\leq a<b\leq 2N$. The entry $ij$ of each element 
of the basis is defined as follows. For $b\neq N+a$,
\beq\label{eq:eij_2AS_dd-dd}
    e^{(ab)}_{ij} \equiv \frac{1}{\sqrt{2}} \left( \delta_{aj}\delta_{bi} - \delta_{ai}\delta_{bj} \right) \; ,
\eeq
while for $b=N+a$ and $2\leq a \leq N$,
\beq\label{eq:eij_2AS_sqrtjjp}
    e^{(ab)}_{i,i+N} = -e^{(ab)}_{i+N,i} \equiv 
    \begin{cases}
        \frac{1}{\sqrt{2a(a-1)}} \; , \;\;\; \text{for} \;\; i<a \,, \\
        \frac{1-a}{\sqrt{2a(a-1)}} \; , \;\;\; \text{for} \;\; i=a \,.
    \end{cases}
\eeq
It is easy to verify that each element of this basis satisfies 
the $\Omega$-traceless condition $\Tr (e^{(ab)} \Omega) = 0 $, where the symplectic matrix $\Omega$ is defined in Eq.~(\ref{eq:Omega_matrix}).

Finally, we impose periodic boundary conditions on the lattice for the link variables, while for the fermions we impose
periodic boundary conditions along the space-like directions, and anti-periodic boundary conditions along the time-like direction.

\section{Numerical Implementation: Grid}
\label{Sec:grid}

Our numerical studies are performed using Grid~\cite{Boyle:2015tjk,Boyle:2016lbp,Yamaguchi:2022feu}: a high level, 
architecture-independent, $\textrm{C++}$ software library for 
lattice gauge theories. 
The portability of its single source-code across 
the many architectures that characterise the exascale platform landscape 
makes it an ideal tool for a long-term computational strategy. 
Grid has already been used to study theories based on $SU(N_c)$ gauge 
groups with $N_c\geq 3$, and fermions in multiple
representations~\cite{Cossu:2019hse, DelDebbio:2022qgu}.
In this section, we describe the changes that have been implemented in 
Grid in order to enable the sampling of $Sp(2N)$ gauge field configurations.
With the aim of including dynamical fermions in future explorations of $Sp(2N)$ 
gauge theories, we focused our efforts\footnote{
An implementation of the Cabibbo-Marinari method~\cite{Cabibbo:1982zn} for pure gauge
theories would be useful to explore general $Sp(2N)$ theories and extrapolate to the 
large-$N_c$ limit. We postpone this task to future work.
} on the Hybrid Monte Carlo (HMC) algorithm and on its variation, the Rational HMC (RHMC), used whenever the number of fermion species is odd. 

The (R)HMC algorithms generate a Markov chain of gauge configurations
distributed as required by the lattice action described in Sect.~\ref{Sec:lattice}. The ideas underpinning these two algorithms
can be summarized as follows.
Firstly, bosonic degrees of freedom $\phi$ and 
$\phi^\dag$, known as pseudofermions, are introduced replacing a generic number $n_f$ of fermions. Powers of the determinant
of the hermitian Dirac operator, $Q^R_m=\gamma_5 D^R_m$, 
in representation $R$ can then be expressed as
\begin{equation}\label{eq:pf}
(\det{D^R_m})^{n_f} = (\det{Q^R_m})^{n_f} = \int \mathcal{D}\phi \mathcal{D}\phi^\dag
e^{-a^4 \sum_x \phi^\dag(x) (Q^2_m)^{-n_f/2} \phi(x)}~,
\end{equation}
where flavor and color indices of $\phi$ and $\phi^\dag$ have been
suppressed for simplicity. For odd values of $n_f$, the rational approximation
is used to compute odd powers of the determinant above, resulting in the RHMC.

Second, a fictitious classical system is defined, with canonical 
coordinates given by  the elementary links and Lie-algebra-valued 
conjugate momenta $\pi(x,\,\mu)=\pi^a(x,\,\mu)\,t^{a}$, where 
$t^{a}$ are the
generators of the $\mathfrak{sp}(2N)$ algebra in the fundamental 
representation. The fictitious hamiltonian is 
\begin{equation}
H=\frac{1}{2}\sum_{x,\mu,a} \pi^a(x,\,\mu) \pi^a (x,\,\mu) + H_g + H_f\,,
\end{equation}
where $H_g=S_g$ and $H_f=S_f$. 
The molecular dynamics (MD) evolution in fictitious time $\tau$ 
is dictated by
\begin{equation}\label{eq:MDev}
\frac{\mathrm{d} U_\mu(x) }{\mathrm{d}\tau} = \pi(x,\,\mu) U_\mu(x)~,\quad
\frac{\mathrm{d} \pi(x,\,\mu)}{\mathrm{d}\tau}= F(x,\mu)\,,
\end{equation}
where $F(x,\mu)$, known as the HMC force, is defined on the Lie algebra
$\mathfrak{sp}(2N)$,
and can be expressed as $F(x,\,\mu) = F_g(x,\,\mu) + F_f(x,\,\mu)$.
The detailed form for $F_g(x,\,\mu)$ and $F_f(x,\,\mu)$~\cite{DelDebbio:2008zf}, 
the gauge and fermion force can be found in Section IIIA of Ref.~\cite{DelDebbio:2008zf}. \\
Numerical integration of the MD equations thus leads to a 
new configuration of the gauge field, which is then accepted or 
rejected according to a Metropolis test. 
The update process can hence be described 
as follows:
\begin{itemize}
\item pseudofermions distributed 
according to the integrand in Eq.~\eqref{eq:pf} are generated with 
the Heat Bath algorithm, 
\item starting with Gaussian random 
conjugate momenta, the MD equations in Eqs.~\eqref{eq:MDev} are integrated
numerically,
\item the resulting gauge configuration is accepted or 
rejected by a Metropolis test. 
\end{itemize}

In this section we provide details on the implementation of the
operations listed above, focusing on the alterations made to the pre-existing
structure of the code designed for $SU(N_c)$ gauge theories. We then describe 
and carry out three types of technical checks, following 
Ref.~\cite{DelDebbio:2008zf}. 
We test the behaviour of the HMC and RHMC 
algorithms. We produce illustrative examples of the behaviour of the 
molecular dynamics (MD). Finally, we carry out a 
comparison between HMC and RHMC algorithms. The purpose of these tests is 
to verify that the dynamics is implemented correctly.

       \begin{figure}[t!]
           \begin{center}
        \includegraphics[width=0.6\linewidth]{ 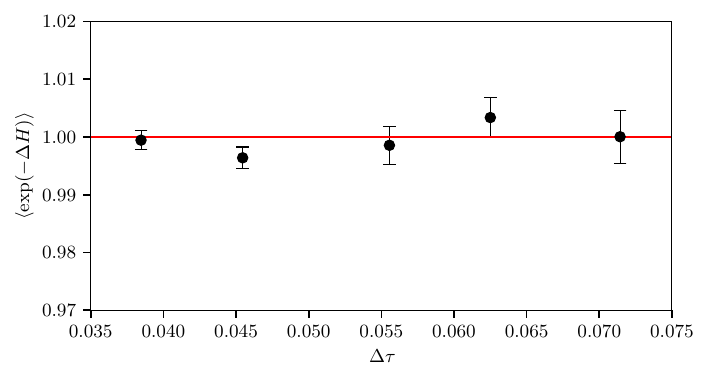}
        \caption{Test of Creutz equality, $\langle\exp(-\Delta H)\rangle = 1$; dependence of $\langle\exp(-\Delta H)\rangle$ on the time–step $\Delta \tau$  in the MD integration, for $N=2$, $N_{\rm f}=0$, and $N_{\rm as}=4$. The relevant parameters of this study are the trajectory length $\tau = 1$, number of steps $n_{\rm steps} = 14, 16, 18, 22, 26$ ($\Delta \tau=\tau/n_{\rm steps}$), for an ensemble with lattice volume $\tilde V/a^4=8^4$, 
        $\beta = 6.8$, and $am_0^{\rm as}  = -0.6$.}\label{Fig:Creutz}
           \end{center}
         \end{figure}

\subsection{Software development}
\label{Sec:software}

As in the case for the pre-existing routines handling the theories with gauge group $SU(N_c)$, 
our implementation of $Sp(2N)$ allows for a generic number of colors.
The starting point of the MD is the generation of random Lie-algebra-valued 
conjugate momenta. The generators of the $\mathfrak{sp}(2N)$ Lie Algebra in 
the fundamental representation, as they appear in Grid, are provided by 
the relations described in Appendix~\ref{app:generators}, where conventions 
for their normalisation are also established. Generators in higher 
representations of the gauge group can be derived 
from the fundamental ones~\cite{DelDebbio:2008zf, Cossu:2019hse}. 
In particular, the generators of the algebra of $Sp(2N)$ in the antisymmetric
representation can be obtained from the definition in Eq.~(\ref{eq:upgrade_Uf_to_Uas}), by
Taylor expanding to first order around the unit transformation,
\beq\label{eq:upgrade_Tf_to_Tas}
  (t_{\mathrm{as}}^{a})_{(ab)(cd)} =  
  \Tr  \left( e^{(ab)\,T} t_{\mathrm{f}}^{a} e^{(cd)} 
  +  e^{(ab)\,T} e^{(cd)} t_{\mathrm{f}}^{a\,T} \right) \; .
\eeq

       \begin{figure}[t!]
         \begin{center}
        \includegraphics[width=0.6\linewidth]{ 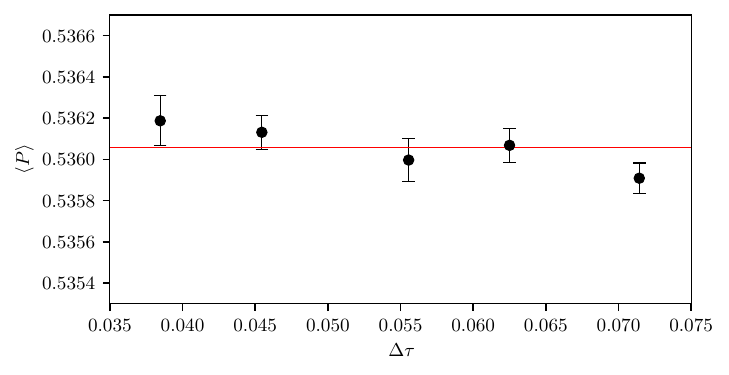}
        \caption{Test of  independence of the plaquette on the time–step $\Delta \tau$
         used for the MD integration, for $N=2$, 
        $N_{\rm f}=0$, and $N_{\rm as}=4$.
The relevant parameters of this study are the  trajectory length $\tau = 1$,  number of steps  $n_{\rm steps} = 14, 16, 18, 22, 26$,  $\Delta \tau=\tau/n_{\rm steps}$, for an ensemble with lattice volume $\tilde V/a^4=8^4$, 
        $\beta = 6.8$, and $am_0^{\rm as}  = -0.6$. The horizontal line corresponds to the plaquette value obtained averaging over trajectories having different a number of step values, $n_{\rm steps}$.}\label{Fig:plaquette}
            \end{center}
         \end{figure}

In the numerical integration of Eq.~\eqref{eq:MDev},
it is required to project the HMC force on the Lie algebra 
of the gauge group.
In Grid, the embedding of the force-projection within the 
integrator requires the forces to be
anti-hermitian. Hence, a projection operation to the matrices 
of the algebra $\mathfrak{sp}(2N)$
must be defined. This can be done in analogy with the 
projection to $\mathfrak{su}(N_c)$, defined for a generic matrix $M$ as
\begin{equation}\label{eq:SUn_TaProj}
    P_{\text{tr}} P_{\text{aH}} M \; , 
\end{equation}
where $P_{\text{tr}} M \equiv M - \mathbb{1}_{N_c} \mathrm{Tr}(M)/N_c$ and
$P_{\text{aH}} M \equiv (M-M^\dagger)/2$ are the projectors to its 
traceless and to its anti-hermitian parts, respectively.
For $\mathfrak{sp}(2N)$, the projection is instead defined as,
\begin{equation}
    P_{\text{aH}} P^{-}_{\text{Sp}} P_{\text{tr}}\,  M \; ,
\end{equation}
where
\begin{equation}
    P^{\pm}_{\text{Sp}} M \equiv \frac{M \pm \Omega M^* \Omega}{2} \;.
\end{equation}
Notice that $P^{-}_{\text{Sp}}$ returns an anti-hermitian matrix, while 
$P^{+}_{\text{Sp}}$ projects on a space of hermitian matrices.

The resympleticisation of gauge links to the $Sp(2N)$ group manifold has 
also been implemented in Grid. The algorithm~\cite{Bennett:2017kga} 
is a modification of the Gram-Schmidt process designed to take into account
the condition in Eq.~\eqref{eq:UOmega_equals_OmegaUstar}. 
After normalising the first column of the matrix $U$, 
the $(N+1)$-th column is set to
 \begin{equation}\label{eq:group_columns}
    \mathrm{col}(U)_{j+N} = -\Omega \, \mathrm{col}(U)_j^* \;.
\end{equation}
The second column is then obtained by orthonormalisation with respect to 
both the first and the $N+1$-th column. An iteration of this process
leads to a $Sp(2N)$ matrix. This procedure, performed after every update, prevents the gauge fields from drifting away from the group manifold due to the finite precision of the simulation.

       \begin{figure}[b]
           \begin{center}
        \includegraphics[width=0.6\linewidth]{ 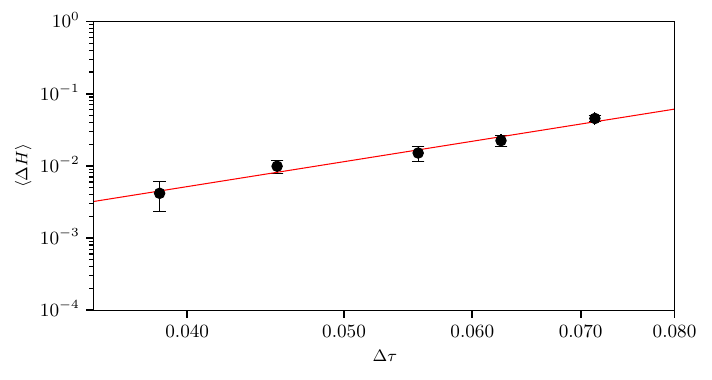}
         \caption{Dependence of $\langle \Delta H \rangle$ on the time-step,  $\Delta \tau$, used for the MD integration, for $N=2$, 
        $N_{\rm f}=0$, and $N_{\rm as}=4$.
 The expectation value $\langle \Delta H \rangle$ is proportional to $(\Delta\tau)^4$, consistently with the use of a second-order integrator. The plot is shown in log-log scale. The relevant parameters of this study are the trajectory length $\tau = 1$,  number of steps  $n_{steps} = 14, 16, 18, 22, 26$  ($\Delta \tau=\tau/n_{\textnormal{steps}}$), for an ensemble with lattice volume $\tilde V/a^4=8^4$, 
        $\beta = 6.8$, and $am_0 = -0.6$.}\label{Fig:DH}
             \end{center}
          \end{figure}

\subsection{Basic tests of the algorithm}
\label{Sec:algorithm}

In this subsection, we follow closely Sects.~III and~IV of Ref.~\cite{DelDebbio:2008zf}.  
Our MD evolution is implemented using a second-order Omelyan integrator~\cite{Takaishi:2005tz}. However, in this work,
the inversion of the fermion matrix is treated without preconditioning~\cite{DeGrand:1990dk, Bennett:2022yfa}. 

\begin{figure}[t]
           \begin{center}
        \includegraphics[width=0.6\linewidth]{ 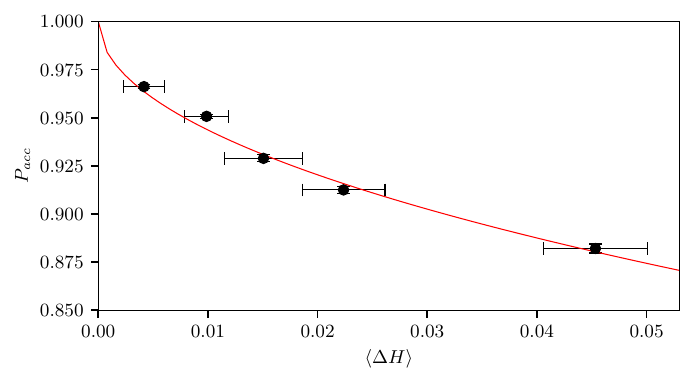}
      \caption{Test of the relation between acceptance probability and $\Delta H$, for $N=2$, 
        $N_{\rm f}=0$, and $N_{\rm as}=4$. The expected behaviour $P_{acc} = \hbox{erfc}(\sqrt{\Delta H}/2)$ is supported. The relevant parameters of this study are the  trajectory length $\tau = 1$,  number of steps  $n_{\rm steps} = 14, 16, 18, 22, 26$  ($\Delta \tau=\tau/n_{\rm steps}$), for an ensemble with lattice volume $\tilde V/a^4=8^4$, 
        $\beta = 6.8$, and $am_0 = -0.6$.}\label{Fig:acceptance}
          \end{center}
       \end{figure}

\begin{figure}[b!]
          \begin{center}
        \includegraphics[width=0.7\linewidth]{ 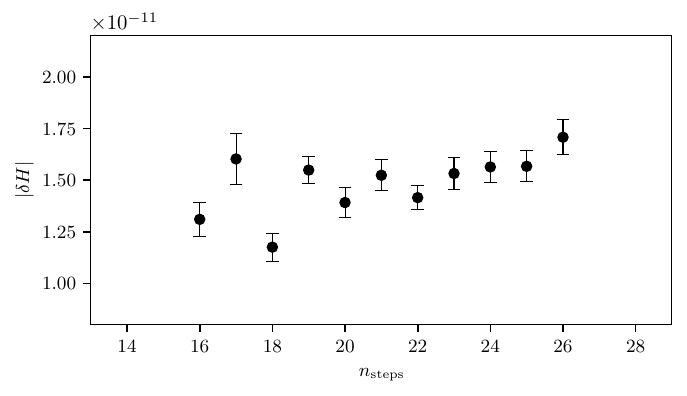}
      \caption{Reversibility test, showing $|\delta H|$ for various choices of $\Delta \tau$, 
      for $N=2$, 
        $N_{\rm f}=0$, and $N_{\rm as}=4$. 
      The relevant parameters of this study are the trajectory length $\tau = 1$,  number of steps  $n_{\rm steps} \in [16, 26]$  ($\Delta \tau=\tau/n_{\rm steps}$), for an ensemble with lattice volume $\tilde V/a^4=8^4$, 
        $\beta = 6.8$, and $am_0 = -0.6$.}\label{Fig:reversibility}
          \end{center}
       \end{figure}

We now restrict attention to the theory with $N=2$, $N_{\rm f}=0$, and $N_{\rm as}=4$, and perform a set of preliminary checks on the algorithms we use. We present the results in Figs.~\ref{Fig:Creutz},~\ref{Fig:plaquette},~\ref{Fig:DH},~\ref{Fig:acceptance}, and~\ref{Fig:reversibility}, obtained, for convenience, setting the lattice parameters to $\beta=6.8$, and $a m_0=-0.6$, on an isotropic lattice with volume $\tilde{V}=(8a)^4$.

The first test pertains to Creutz equality~\cite{Creutz:1988wv}: by measuring the difference in Hamiltonian, $\Delta H$, evaluated before and after the MD evolution, one should find that 
\beqs
\left\langle \exp\left(-\frac{}{}\Delta H\right)\right\rangle&=&1\,.
\eeqs
This is supported by our numerical results: Fig.~\ref{Fig:Creutz} shows the value of 
$\left\langle \exp\left(-\frac{}{}\Delta H\right)\right\rangle$ for five different choices of the time-step used in the 
MD integration, with $\Delta \tau=\tau/n_{steps}$, and the choice $\tau=1$. 
The numerical results are obtained by considering a thermalised ensemble consisting 
of $3400$ trajectories, that we find has integrated auto-correlation time $\tau_c=6.1(2)$,
measured using the Madras-Sokal windowing process~\cite{Madras:1988ei}.
A closely related test is shown in Fig.~\ref{Fig:plaquette}:  the value of the ensemble average of the plaquette
 is independent of $\Delta \tau$.

A third test pertains to the dependence of $\langle \Delta H\rangle$ on $\Delta \tau$, which for a second-order
integrator is supposed to scale as $\langle \Delta H\rangle \propto (\Delta \tau)^4$~\cite{Takaishi:1999bi}.
In Fig.~\ref{Fig:DH} we show our measurements, together with the result of a best-fit to the curve $\log \langle \Delta H \rangle = {\cal K}_1 \,\log (\Delta \tau) + {\cal K}_2$, with ${\cal K}_1 = 3.6(4)$ determined by minimising a simple $\chi^2$. We find good agreement, as quantified by the value of the reduced $\chi^2/N_{\rm d.o.f.}=0.6$, and ${\cal K}_1$ is compatible to $4$. A closely related test is displayed in Fig.~\ref{Fig:acceptance}, confirming the prediction that the 
acceptance probability of the algorithm, $P_{\rm acc}$, obeys the relation~\cite{Gupta:1990ka}: 
\beqs
P_{\rm acc}&=&{\rm erfc}\left(\frac{1}{2}\sqrt{\left\langle\Delta\frac{}{} H\right\rangle}\right)\,.
\eeqs

   \begin{figure}[t]
 \begin{center}
        \includegraphics{ 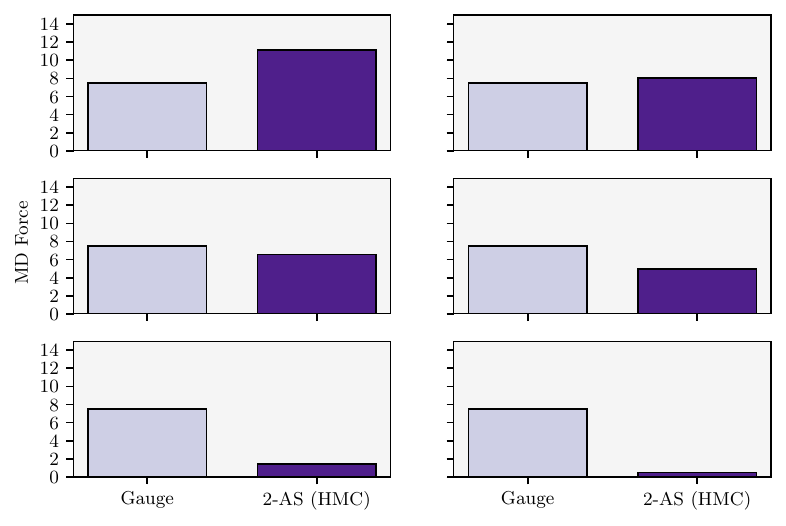}
            \end{center}
      \caption{Field contribution to the MD force for the theory with $N=2$, $N_{\rm f}=0$, and $N_{\rm as}=4$, on isotropic lattice with $\tilde V=(8a)^4$, and lattice coupling  $\beta = 6.8$.
The two blocks are respectively indicating the gauge (light shading, left) and the fermion (dark shading, right) contribution, the latter computed with the HMC algorithm. Fermion contributions are summed over flavor. The six panels correspond to different choices of bare mass: $am_0^{\rm as} = -0.9,\, -0.1,\,+0.6,\,+1.8,\,+15,\,+50$ (left to right, top to bottom).}
\label{Fig:MD}
       \end{figure}

The final test of this subsection is displayed in Fig.~\ref{Fig:reversibility}.
Following Refs.~\cite{DelDebbio:2008zf,Joo:2000dh}, we also want to ensure that the reversibility of our updates is respected. Reversibility is one of the fundamental properties required in order to pursue a correct HMC update. Our update algorithm, based on leapfrog, is reversible analytically. Yet, when using this algorithm numerically on computers, because of the finite precision, exact reversibility is lost. It is therefore important to verify that implementation of the fundamental steps of the algorithm can be considered as reversible to good approximation, in order to avoid that rounding errors introduce a significant bias in our calculations. One can show that the quantity $|\delta H|$---the average difference of the Hamiltonian evaluated by evolving the MD forward and backward and flipping the momenta at $\tau=1$---doesn't change significantly in our simulations. 
Since the Hamiltonian in these tests is of order $\sim10^6$ and the typical $\delta H\sim10^{-11}$, the results show that the violation of reversibility is consistent with having $|\delta H| / H$ of the order of the numerical accuracy. This is the expected relative precision for double-precision floating-point numbers. Moreover, the violation $|\delta H|$ is independent of $\Delta \tau$.  As a ``microscopic" and related effect, reversibility violations may occur while updating the gauge link variables and momenta updates during the MD evolution. To ensure this doesn't occur we update the gauge links through the exponentiation of the momenta, so that $U (\pi) \, U (-\pi) = 1$. Moreover, thanks to the double-precision nature of the variables we use, the entity of relative violations for momenta results to be within machine-accuracy in our simulations, as in the global reversibility violation case.

\subsection{More about the Molecular Dynamics}
\label{Sec:MP}

\begin{figure}[t]
\begin{center}
\begin{picture}(100,130)
     \put(0,65){ \includegraphics[width=0.60\linewidth]{ 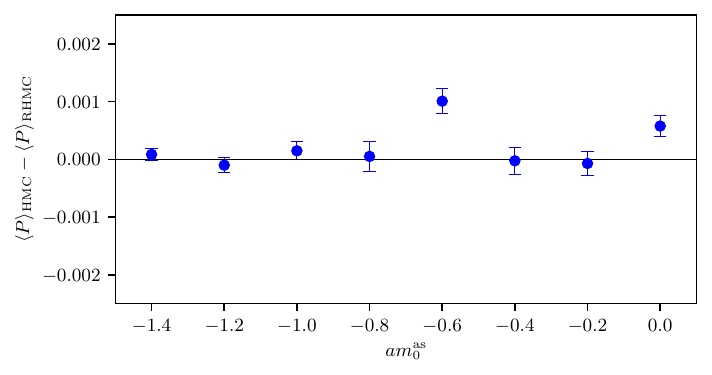}}
      \put(0,0){ \includegraphics[width=0.60\linewidth]{ 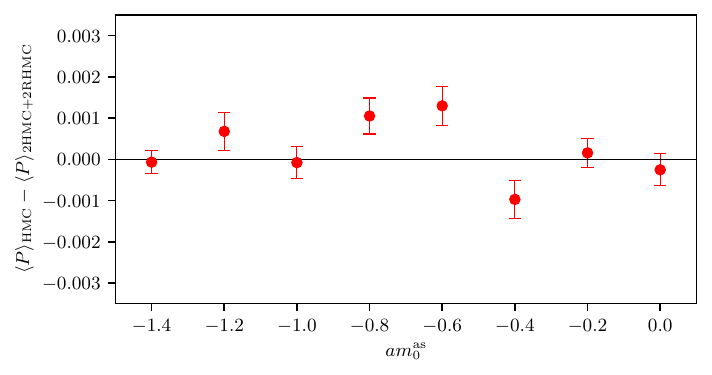}}
      \end{picture}
      \caption{Compatibility between plaquette averages $\langle P \rangle$ obtained with HMC and RHMC algorithms for the theory with $N=2$, $N_{\rm f}=0$, and $N_{\rm as}=4$.  $\langle P \rangle_{\rm HMC}$ is obtained running two couples of fermions with HMC. For $\langle P \rangle_{\rm RHMC}$ (top panel), RHMC was applied individually to each of the fermions.  $\langle P \rangle_{\rm 2HMC + 2RHMC}$ (bottom panel) is obtained running two fermions with HMC, while the other two were run with RHMC. 
      The lattice coupling is $\beta = 6.8$, with the bare mass in the range $-1.4 \leq a m^{\rm as}_0 \leq 0.0$. 
      The lattice is isotropic and has volume $\tilde V = (8a)^4$.}
\label{Fig:RHMC}
\end{center}
\end{figure}

For illustration purposes, we find it useful to monitor the contribution to the MD of the fields, and how this changes as we dial the lattice parameters. We focus on the theory with $N=2$, $N_{\rm f}=0$, and $N_{\rm as}=4$,
and consider a few ensembles with isotropic lattice with $\tilde V=(8a)^4$, and lattice coupling  $\beta = 6.8$,
but vary the mass $a m_0^{\rm as}$.
We show in Fig.~\ref{Fig:MD}
the force,  $F$, as defined in Eq.~(\ref{eq:MDev}), split in its 
contribution from the gauge and fermion dynamics, the latter computed using the HMC for all fermions.
The results are normalised so that the gauge contribution is held constant.
As can be clearly appreciated, for large and positive values of $a m_0^{\rm as}$ the fermions can be neglected,
as for these choices of the mass, one expects to be in the quenched regime. 
When decreasing the mass, the fermion contribution increases. For large, negative values of the Wilson bare mass (close to the chiral limit),
the fermion contribution is even larger than the contribution of the gauge part of the action.

\subsection{Comparing HMC and RHMC}
\label{Sec:HMCvsRHMC}

While in this paper we are mostly interested in the theory with  $N=2$, $N_{\rm f}=0$, and $N_{\rm as}=4$, and hence we can use the HMC algorithm, for the general purpose of identifying the extent of the conformal window in this class of lattice gauge theories it may be necessary to consider also odd numbers of fermions, for which we resort to the RHMC algorithm. The latter relies on a rational approximation in the computation of the fermion force, but the presence of a Metropolis accept-reject step ensures that the algorithm is exact. Thus, a preliminary test must be made to check the consistency of the implementation---as was done for $SU(3)$ theories, see for instance Ref.~\cite{Clark:2003na}.\footnote{We note that to check the correctness of the Remez implementation, one could in principle use any function of an arbitrary matrix $M$. In particular, choosing diagonal matrices would make the comparison straightforward. Grid makes use of this methodology in its test suite.} To gauge whether the numerical implementation is working at the desired level of accuracy and precision, we performed the exercise leading to Fig.~\ref{Fig:RHMC}.
We computed the average plaquette, $\langle P\rangle$, where $P$ is defined as
\beq
P \equiv \dfrac{a^4}{6\tilde{V}}\sum_x \sum_{\mu<\nu} \left[ \frac{1}{2N}{\rm Re}\, {\Tr}\, {\cal P}_{\mu\nu}(x) \right]
\eeq
for ensembles having lattice
volume $\tilde V = (8a)^4$ and coupling $\beta = 6.8$, for a few representative choices of the bare mass $-1.4 \leq a m^{\rm as}_0 \leq 0.0$.
We repeated this exercise three times: at first, we treated all fermions with the HMC, then we treated them all with the RHMC, and finally we used a mixed strategy, treating two fermions with the HMC, and two with the RHMC.
We display, in the two plots in the figure, the differences of the second and third approaches to the first one, respectively.
For most of the data, the differences are compatible with zero within the statistical uncertainties. More generally, given the number of observables, the probability to find a deviation larger than $3\sigma$ from the null value results to be $\sim 12\%$.

   \begin{figure}
 \begin{center}
        \includegraphics{ 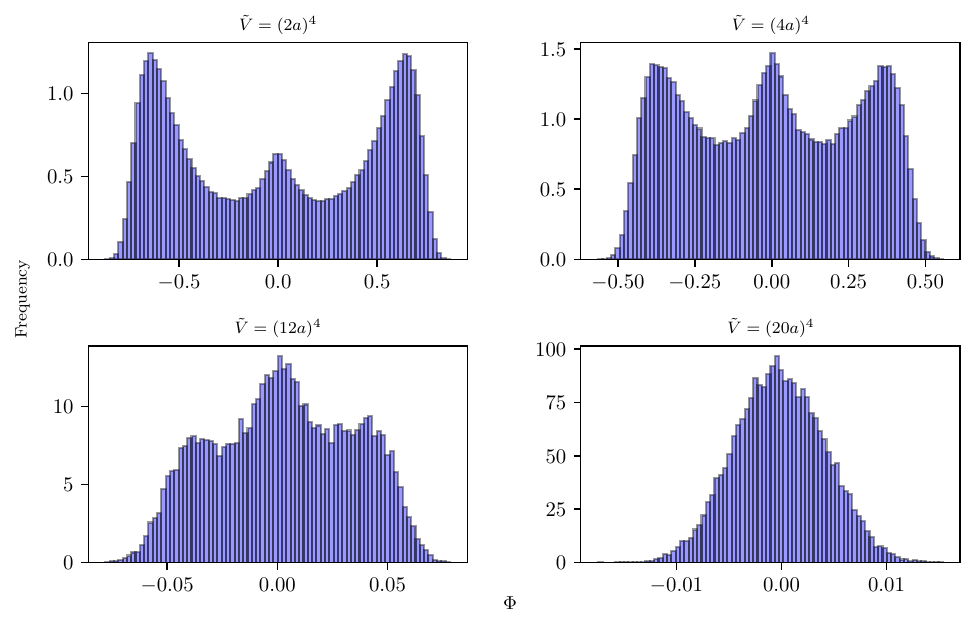}
            \end{center}
      \caption{Study of finite-size effects on the lattice, for the $Sp(4)$ Yang-Mills theory. The histograms depict the distribution of (real) Polyakov loops for ensembles with $\beta =9.0$ and four choices of space-time volume: 
       $\tilde{V} = (2a)^4,\, (4a)^4,\, (12a)^4,\, (20a)^4$. The histograms' areas are normalised to $1$.}
      \label{Fig:Polyakov}
       \end{figure}
       
\section{The $N=2$ lattice Yang-Mills theory}
\label{Sec:2}

In this section, we start to analyse the physics of the $Sp(4)$ theory of interest.
We begin from the pure Yang-Mills dynamics, with $N_{\rm f}=0=N_{\rm as}$.
We verify that centre symmetry, $\left({\mathbb Z}_2\right)^4$, is broken at small volumes, but restored at large volumes,
by looking at the (real) Polyakov loop,  in a way that is reminiscent of Ref.~\cite{Athenodorou:2014eua}.
Following Ref.~\cite{Cossu:2019hse}, we then consider the spectrum of the Dirac operator in the quenched approximation, both for fundamental and 2-index antisymmetric fermions, to verify the symmetry-breaking pattern expected from random matrix theory.

The results for the first of these tests are shown in Fig.~\ref{Fig:Polyakov}. At a coupling $\beta=9.0$, we generate four ensembles in the pure $Sp(4)$ theory, at different values of the space-time volume,
 $\tilde{V}= (2a)^4,\, (4a)^4,\, (12a)^4,\, (20a)^4$. For each configuration, we compute the spatial averaged (real) Polyakov loop, defined as
\beqs
\Phi&\equiv & 
\frac{1}{N_c N_s^3} \sum_{\vec{x}} \Tr \left(\prod_{t=0}^{t=N_t-1} U_0(t,\vec{x})\right)\,,
\eeqs
where $U_0(t,\vec{x})$ is the time-like link variable. 
For our current purposes, we choose the lattice to be isotropic in all four directions, $N_t=N_s=L/a$. 
For each ensemble, we display the frequency histogram of the values of $\Phi$.
The expectation is that the zero-temperature $Sp(4)$ lattice theory should preserve the $\left({\mathbb Z}_2\right)^4$ symmetry of the centre of the group in four Euclidean space-time dimensions.
This is indeed the case for sufficiently large volumes, as shown by the bottom-right panel of Fig.~\ref{Fig:Polyakov}, 
for which $N_t=N_s=20$,
that exhibits a Gaussian distribution centred at the origin.
But for small enough lattice volumes, this expectation is violated. This is visible in the other three panels in Fig.~\ref{Fig:Polyakov}, 
in which the distribution 
is non-Gaussian, and two other peaks emerge. In the extreme case of $N_s=N_t=2$, the two peaks at finite value of $\Phi$
dominate the distribution, which is otherwise symmetrical around zero. Interestingly, Polyakov loops can be  used also to perform the more physical study of the finite-temperature confinement/deconfinement phase transition. In this case, one would consider  $N_t \neq N_s$, vary the coupling $\beta$ to identify the transition temperature, and then perform continuum and infinite volume extrapolations. The characterisation of the deconfinement phase transition is of interest for both theoretical as well as phenomenological reasons---see Ref.~\cite{Holland:2003kg} and the review in Ref.~\cite{Bennett:2023wjw}---but requires a dedicated, extensive programme of numerical work, and possibly cutting-edge new
technology to address some of the difficulties faced by conventional Monte Carlo sampling methods (see, e.g., the discussions in Refs.~\cite{Borsanyi:2022xml} and~\cite{Lucini:2023irm}), while
the simpler analysis performed here suffices for the more modest purposes of this paper. 

   \begin{figure}
 \begin{center}
        \includegraphics[width=0.88\linewidth]{ 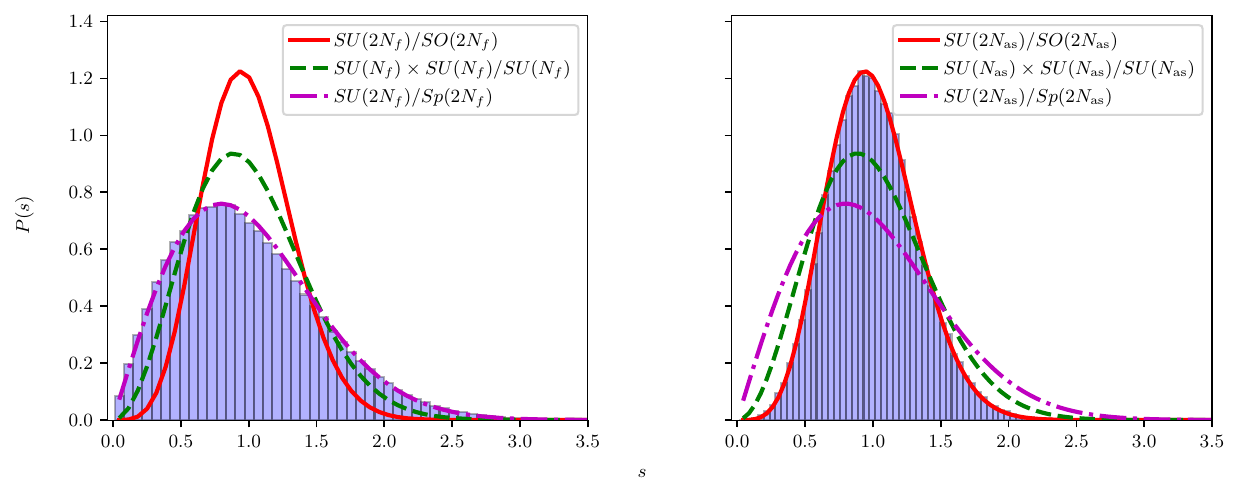}
            \end{center}
      \caption{Distribution of the folded density of spacing between subsequent eigenvalues of the hermitian Dirac-Wilson operator $Q_m = \gamma_5 D_m$, and comparison with predictions from chRMT, computed in the quenched approximation, with ensembles having $\beta = 8.0$, $am_0 = -0.2$, and lattice volume $\tilde V=(4a)^4$, in the $Sp(4)$ theory. The left panel shows the case of fermions transforming in the fundamental representation, and the right is for fermions in the 2-index antisymmetric one.}
      \label{Fig:HDW}
       \end{figure}

Ensembles of gauge configurations without dynamical fermions can also be used to verify that our implementation of the Dirac operators is correct. To this purpose, following Ref.~\cite{Cossu:2019hse} (and Ref.~\cite{Bennett:2022yfa}), we consider the theory with quenched fermions in either the fundamental or 2-index antisymmetric representation, and compute the spectrum of eigenvalues of the hermitian Wilson-Dirac operator $Q_m = \gamma_5 D_m$. The numbers of configurations are $N_{\rm{conf, f}} = 88$ and $N_{\rm{conf, as}} = 47$, while the number of eigenvalues in each configuration
used is $3696$ for fundamental fermions and $5120$ for antisymmetric fermions. Then, we compute the distribution of the folded density of spacing, $P(s)$. Finally, we compare the results to the exact predictions of chiral Random Matrix Theory (chRMT)~\cite{Verbaarschot:1994qf, Verbaarschot:2000dy}. 
Because the spectrum captures the properties of the theory, in particular the pattern of chiral symmetry breaking~\cite{Banks:1979yr}, the distribution $P(s)$ differs, depending on the symmetry-breaking pattern predicted.
The folded density of spacing is
\beq
\label{eq:folded_density}
P(s) = N_{\tilde{\beta}} s^{\tilde{\beta}} \exp\left(-c_{\tilde{\beta}} s^2\right)\,,  \quad  \hbox{where} \quad N_{\tilde{\beta}} = 2 \dfrac{\Gamma^{\tilde{\beta}+1}\left( \frac{\tilde{\beta}}{2} + 1  \right)}{\Gamma^{\tilde{\beta}+2}\left( \frac{\tilde{\beta}+1}{2}  \right)} , \, c_{\tilde{\beta}} = \dfrac{\Gamma^{2}\left( \frac{\tilde{\beta}}{2} + 1  \right)}{\Gamma^{2}\left( \frac{\tilde{\beta}+1}{2}  \right)}\,,
\eeq
where $\tilde{\beta}$ is the Dyson index. This index can take three different values: $\tilde{\beta} = 2$ corresponds to the symmetry breaking pattern $SU(N_f) \times SU(N_f) \to SU(N_f)$, $\tilde{\beta} = 1$ to $SU(2N_{f}) \to Sp(2N_{f})$, and $\tilde{\beta} = 4$ to $SU(2N_{f}) \to SO(2N_{f})$. The latter two are the cases we are interested in, corresponding to fundamental and 2-index antisymmetric fermions for the symplectic theory. 

In order to make a comparison with the chRMT prediction in Eq.~(\ref{eq:folded_density}), we compute the eigenvalues of $Q_m$ for $N_{\rm conf}$ configurations. This process yields a set of eigenvalues $\lambda^{(c)}_i$ with $c = 1, \cdots, N_{\rm conf}$. The eigenvalues are arranged  in one long list, in which $\lambda^{(c)}_i$ are ordered in ascending order. Any degeneracy that is present in the 2-antisymmetric case is discarded. Then, for each $c = 1, \cdots, N_{\rm conf}$, a new list of values is produced, that contains $n^{(c)}_i$, the positive integer position of the eigenvalue $\lambda^{(c)}_i$ in the long list ordered in ascending order, instead of $\lambda^{(c)}_i$. The density of spacing, $s$, is replaced by the expression
\beq
s = \dfrac{n^{(c)}_{i+1} - n^{(c)}_{i}}{\mathcal{N}}\,.
\eeq
The constant $\mathcal{N}$ is defined so that the density of spacing has unit average over the whole ensemble, $\langle s \rangle = 1$. Finally, the (discretised) unfolded density of spacings, $P(s)$, is obtained by binning numerical results for $s$ and normalising it.

   \begin{figure}[t]
 \begin{center}
        \includegraphics[width=0.64\linewidth]{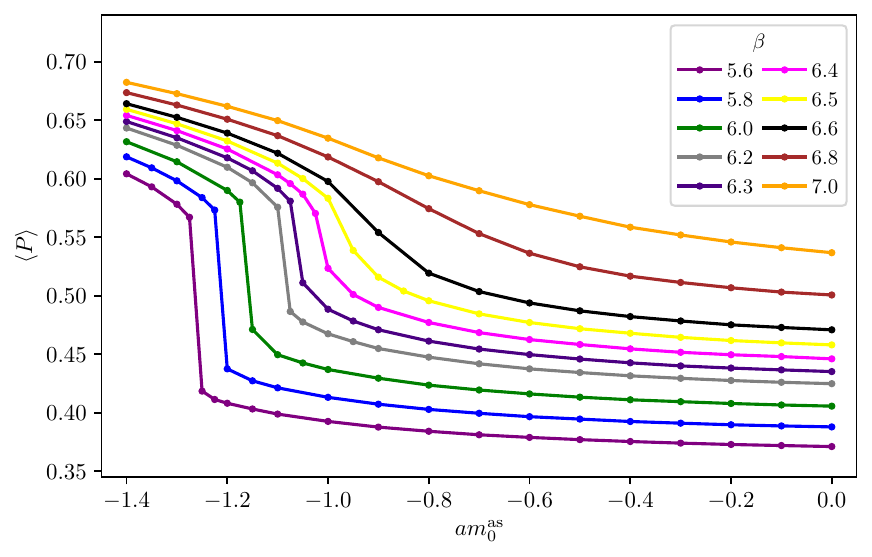}
            \end{center}
      \caption{Parameter scan of the $Sp(4)$ theory with $N_{\rm as}=4$ fermions transforming  in the 2-index antisymmetric representation, with ensembles generated from a cold start, using the HMC. We show the value of
  the average plaquette, $\langle P \rangle$, as a function of the bare mass, for a few representative values of the coupling.  The lattice size is $\tilde V=(8a)^4$, and each point is obtained by varying the lattice coupling  $\beta = 7.0, 6.8, 6.6, 6.5, 6.4, 6.3, 6.2, 6.0, 5.8, 5.6$ and the bare mass $-1.4 \leq a m_0^{\rm as} \leq 0.0$. }
      \label{Fig:Sp4Scan}
       \end{figure}

In Fig.~\ref{Fig:HDW}, we show an example of the folded distribution of eigenvalues of the Wilson-Dirac operator, computed numerically. As it can be seen, in the case of fermions in the fundamental representation, one finds a distribution that is compatible with the symmetry breaking pattern leading to the coset $SU(2N_{\rm f})/Sp(2N_{\rm f})$. Conversely, for fermions in the 2-index antisymmetric representation,  our numerical results reproduce the prediction associated with the coset $SU(2N_{\rm as})/SO(2N_{\rm as})$. The spectacular agreement with chRMT confirms that there are no inconsistencies in our way of treating fermions. The size of the lattices we have considered has been chosen in order to reduce finite-size effects. These effects, as shown in Ref.~\cite{Bennett:2022yfa}, can become evident in smaller lattices and they lead to discrepancies due to some abnormally large spacings for the smallest and largest eigenvalues. This was interpreted to be an artefact due to the finiteness of the lattice size. We observe that, as in previous studies, the antisymmetric representation already matches the predictions in a $4^4$ volume, while for the fundamental to reproduce the predictions chRMT, we had to remove the $200$ lowest and highest eigenvalues (reducing the number of eigenvalues from $4096$ to $3696$). In this fashion, the differences with chRMT are no longer visible to the naked eye even for lattices with modest volume, $\tilde V=(4a)^4$.

\section{The $N=2$  theories coupled to fermions: bulk phase structure}
\label{Sec:2-4}

In this section, we present our main results for the theory with $N=2$, $N_{\rm f}=0$, and varying number of fermions transforming in the antisymmetric representation, starting from $N_{\rm as}=4$---for which we apply the HMC algorithm.
We performed a coarse scan of the lattice parameter space, to identify phase transitions in the $(\beta,m_0)$ plane,
by studying the average plaquette, $\langle P \rangle$, its hysteresis, and its susceptibility.
We provide an approximate estimate of the upper bound coupling for the bulk phase,  $\beta_{\ast}$, above which there is no bulk phase transition, and hence one can safely perform
lattice numerical calculations at finite lattice spacing, yet confident that the
results can be extrapolated to the appropriate continuum limit.

   \begin{figure}
 \begin{center}
        \includegraphics[width=0.64\linewidth]{ 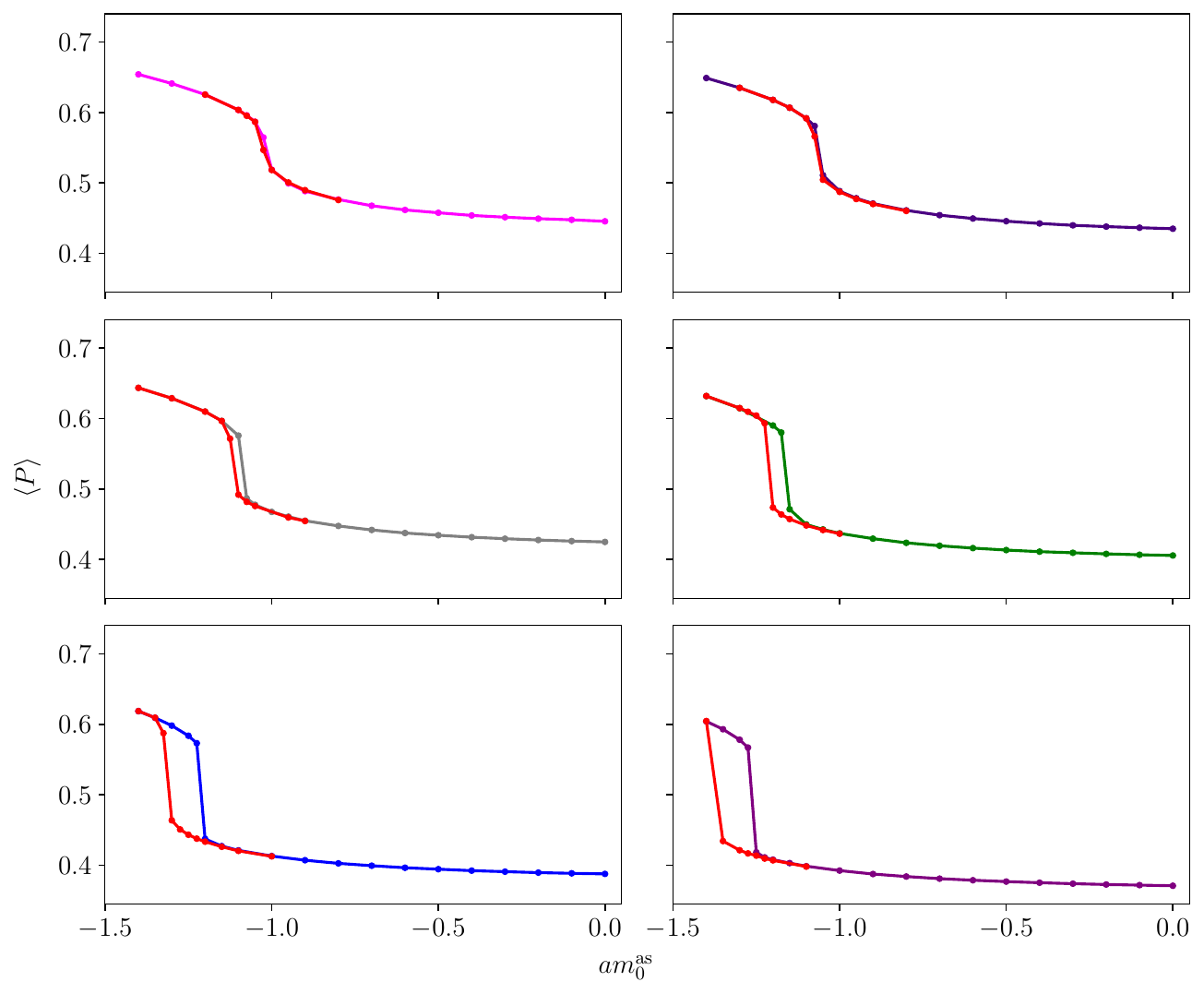}
            \end{center}
      \caption{Hysteresis between hot (red) and cold (other colors) starts for the $Sp(4)$ theory with $N_{\rm as}=4$ fermions in the 2-index antisymmetric representation. The lattice coupling is $\beta = 6.4, 6.3, 6.2, 6.0, 5.8, 5.6$ (left to right, and top to bottom). The lattice size is $\tilde V=(8a)^4$, and each point is obtained by varying  the bare mass $-1.4 \leq a m_0^{\rm as} \leq 0.0$.}
      \label{Fig:hysteresis}
       \end{figure}

Figure~\ref{Fig:Sp4Scan} displays the average plaquette, $\langle P \rangle$, 
 obtained in ensembles generated using a cold start. The lattice size is $\hat V=(8a)^4$, and each point is obtained by varying the lattice coupling  $\beta = 7.0, 6.8, 6.6, 6.5, 6.4, 6.3, 6.2, 6.0, 5.8, 5.6$ and the bare mass $-1.4 \leq a m_0^{\rm as} \leq 0.0$.
 The figure shows that, for small values of $\beta$ and large, negative values of the bare mass, the average plaquette displays an abrupt change at a particular value  $a m_0^{{\rm as}\,\ast}$, while being a smooth, continuous function elsewhere.
 This is a first indication of the existence of a first-order bulk phase transition.

To better understand whether a first-order phase transition is taking place, we study the effect of adopting two different strategies in the generation of the ensembles, repeating it using of thermalised (hot) starts, and redoing the measurements.
Figure~\ref{Fig:hysteresis} shows the comparison of the average plaquette, $\langle P \rangle$, computed for several fixed choices of the coupling $\beta$, while varying the bare mass $-1.4 \leq a m_0^{\rm as} \leq 0.0$. The two curves in the plots represent the behaviour measured in ensembles obtained from a cold and hot start configuration.
The effects of hysteresis are clearly visible for $\beta<6.4$ and are an indication of the presence of a first-order 
 phase transition taking place at a critical value of the bare mass $a m_0^{{\rm as}\,\ast}$.

\begin{figure}
 \begin{center}
        \includegraphics[width=0.80\linewidth]{ 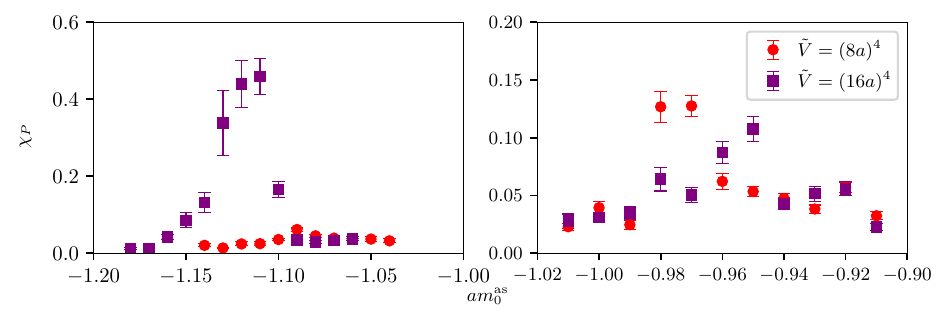}
            \end{center}
      \caption{Plaquette susceptibility, $\chi_P$, in the $Sp(4)$ lattice theory with $N_{\rm as}=4$ fermions in the 2-index antisymmetric representation. We use two values of the lattice size, $\tilde V=(8a)^4$ and $\tilde V=(16a)^4$. The ensembles have 
  $\beta = 6.2$, $-1.18 \leq am_0 \leq -1.04$ (left panel), and  $\beta = 6.5$, $-1.01 \leq am_0 \leq -0.91$ (right panel).}
      \label{Fig:susceptibility}
       \end{figure}

The final test of the nature of the phase transition is shown in Fig.~\ref{Fig:susceptibility}.
For illustration purposes, we choose two values of the coupling for which we have evidence of 
a phase transition ($\beta=6.2$), or of smooth behaviour of $\langle P\rangle$ for all value of $a m_0^{\rm as}$ ($\beta=6.5$), respectively.
We compute the plaquette susceptibility, defined as
\beqs
\chi_P&\equiv& \frac{\tilde{V}}{a^4} \left( \langle P^2\rangle - \left(\langle P\rangle \right)^2 \right) \,,
\eeqs
and compare the numerical results obtained with ensembles having two different volumes, 
$\tilde V=(8a)^4$ and $\tilde V=(16a)^4$.
The results indicate that the peak height scales as the $4$-volume when $\beta$ is small, in which case 
the position of the peak also moves to a different value of $a m_0^{\rm as}$. These are indeed the expected signature of a first order phase transition. For large $\beta$, we observe a shift in values of $am_0^{\rm as}$ and no clear and large change in entity for the peak heights between the two volumes. This is a clear indication of a smooth crossover.
 We hence conclude that, in the theory with $N=2$, $N_{\rm f}=0$, and $N_{\rm as}=4$,
  there is numerical evidence of a line of first-order phase transitions turning into a crossover at $\beta > \beta_{\ast}=6.4$.

\begin{figure}[t] 
\begin{center}
\begin{picture}(170,200)
     \put(0,150){ \includegraphics[width=0.44\linewidth]{ 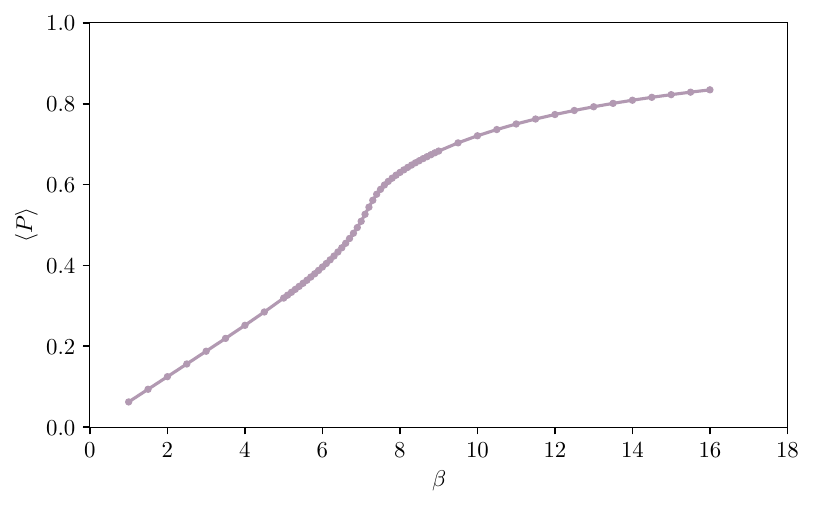}}
     \put(85,150){ \includegraphics[width=0.44\linewidth]{ 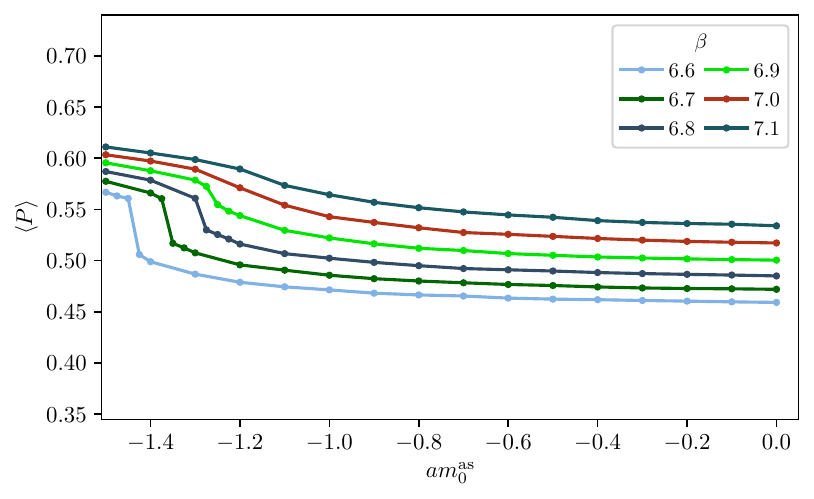}}
          \put(0,100){ \includegraphics[width=0.44\linewidth]{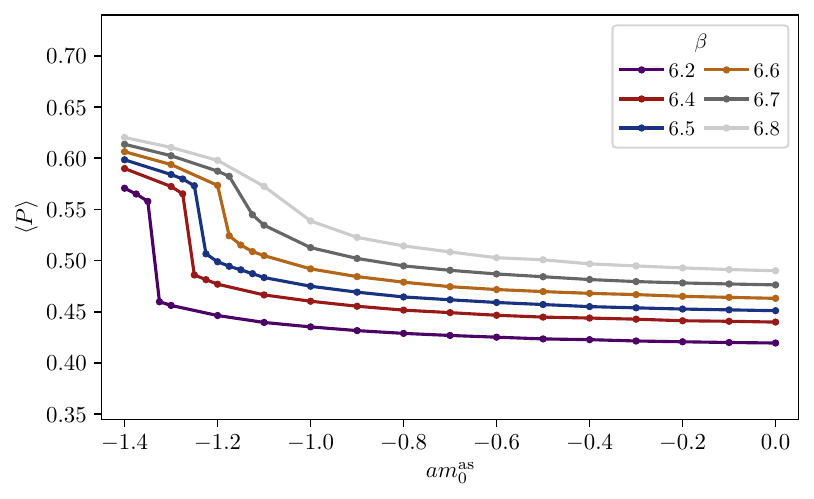}}
     \put(85,100){ \includegraphics[width=0.44\linewidth]{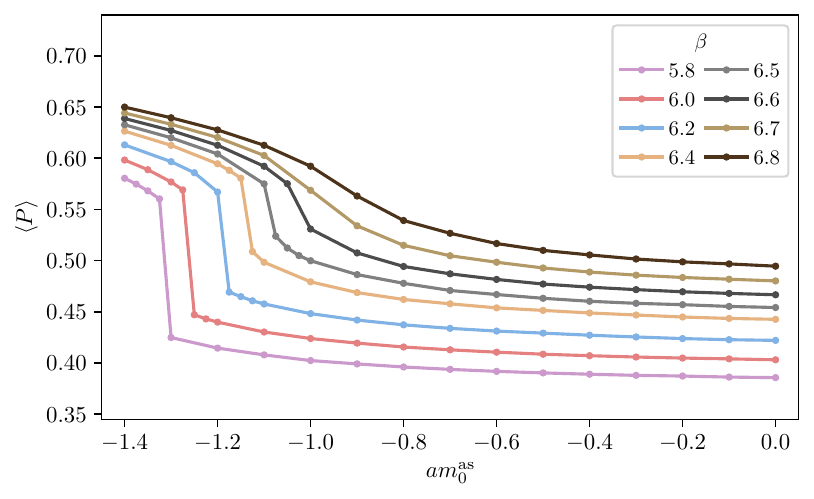}}
          \put(0,50){ \includegraphics[width=0.44\linewidth]{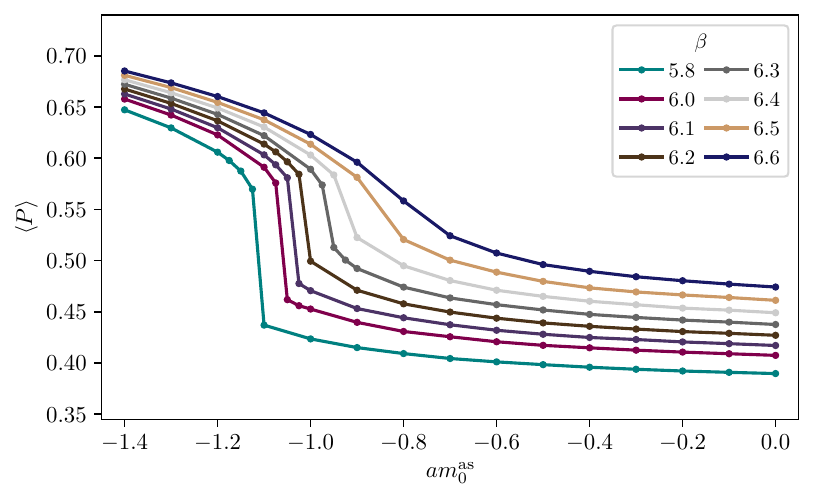}}
     \put(85,50){ \includegraphics[width=0.44\linewidth]{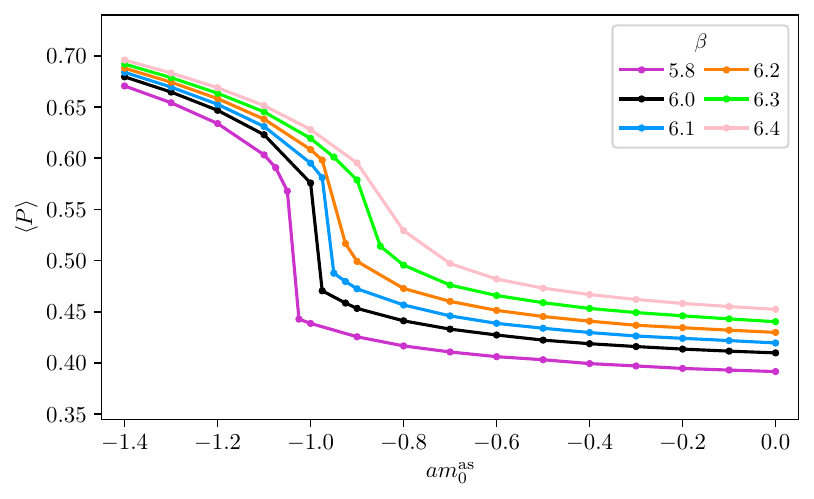}}
     \put(0,0){ \includegraphics[width=0.44\linewidth]{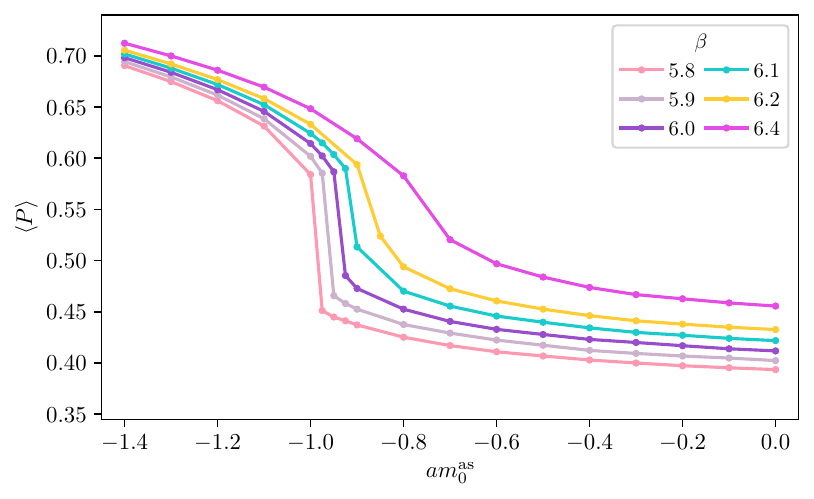}}
     \put(85,0){ \includegraphics[width=0.44\linewidth]{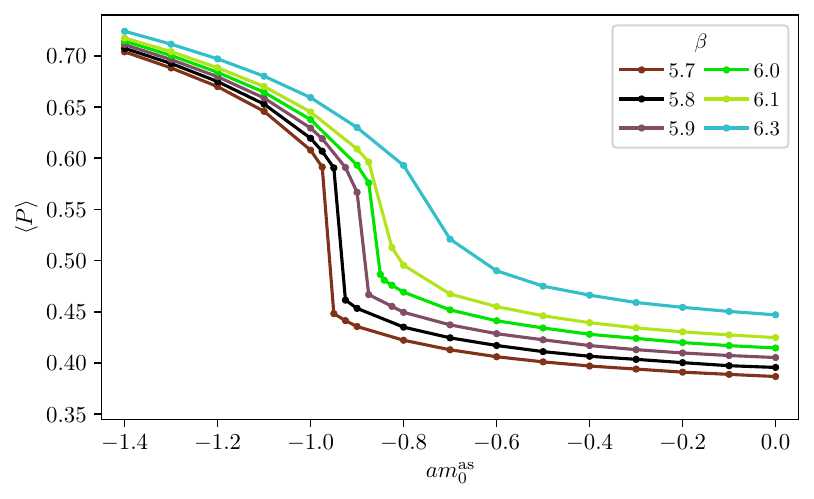}}
     \put(40,190){$N_{\rm as} = 0$}
        \put(125,190){$N_{\rm as} = 1$}
             \put(40,140){$N_{\rm as} = 2$}
                 \put(125,140){$N_{\rm as} = 3$}
                     \put(40,90){$N_{\rm as} = 5$}
                         \put(125,90){$N_{\rm as} = 6$}
                             \put(40,40){$N_{\rm as} = 7$}
                                 \put(125,40){$N_{\rm as} = 8$}
      \end{picture}
      \end{center}
  \caption{Parameter scan in the $Sp(4)$ theory with $N_{\rm as} = 0,\, 1,\, 2,\, 3,\, 5,\, 6,\, 7,\, 8$ (left to right and top to bottom panels) 
  fermions in the 2-index antisymmetric representation, obtained with ensembles generated from a cold start.  For $N_{\rm as} > 0$, we show the value of
  the average plaquette, $\langle P \rangle$, as a function of the bare mass, for a few representative values of the coupling. For pure gauge, we just vary the value of $\beta$.
  All the fermions are treated with the HMC/RHMC algorithms. The lattice size is $\tilde{V} = (8a)^4$ and the base mass is chosen in the range $-1.4 \leq a m_0^{\rm as} \leq 0.0$ for $N_{\rm as} \geq 2$, and $-1.5 \leq a m_0^{\rm as} \leq 0.0$ for $N_{\rm as} = 1$. For the pure gauge theory, the coupling is chosen to be $1.0 \leq \beta \leq 16.0$. For $N_{\rm as} = 1$, we have chosen $\beta = 7.1, 7.0, 6.9, 6.8, 6.7, 6.6$, while for $N_{\rm as} = 2$ we have $\beta = 6.8, 6.7, 6.6, 6.5, 6.4, 6.2$. For $N_{\rm as} = 3$, the coupling is $\beta = 6.8, 6.7, 6.6, 6.5, 6.4, 6.2, 6.0, 5.8$, while for
   $N_{\rm as} = 5$ we've chosen $\beta = 6.6, 6.5, 6.4, 6.3, 6.2, 6.1, 6.0, 5.8$. For $N_{\rm as} = 6$, $\beta = 6.4, 6.3, 6.2, 6.1, 6.0, 5.8$.  For $N_{\rm as} = 7$, $\beta = 6.4, 6.2, 6.1, 6.0, 5.9, 5.8$ and for $N_{\rm as} = 8$, $\beta = 6.3, 6.1, 6.0, 5.9, 5.8, 5.7$.}
\label{Fig:3-5}
\end{figure}
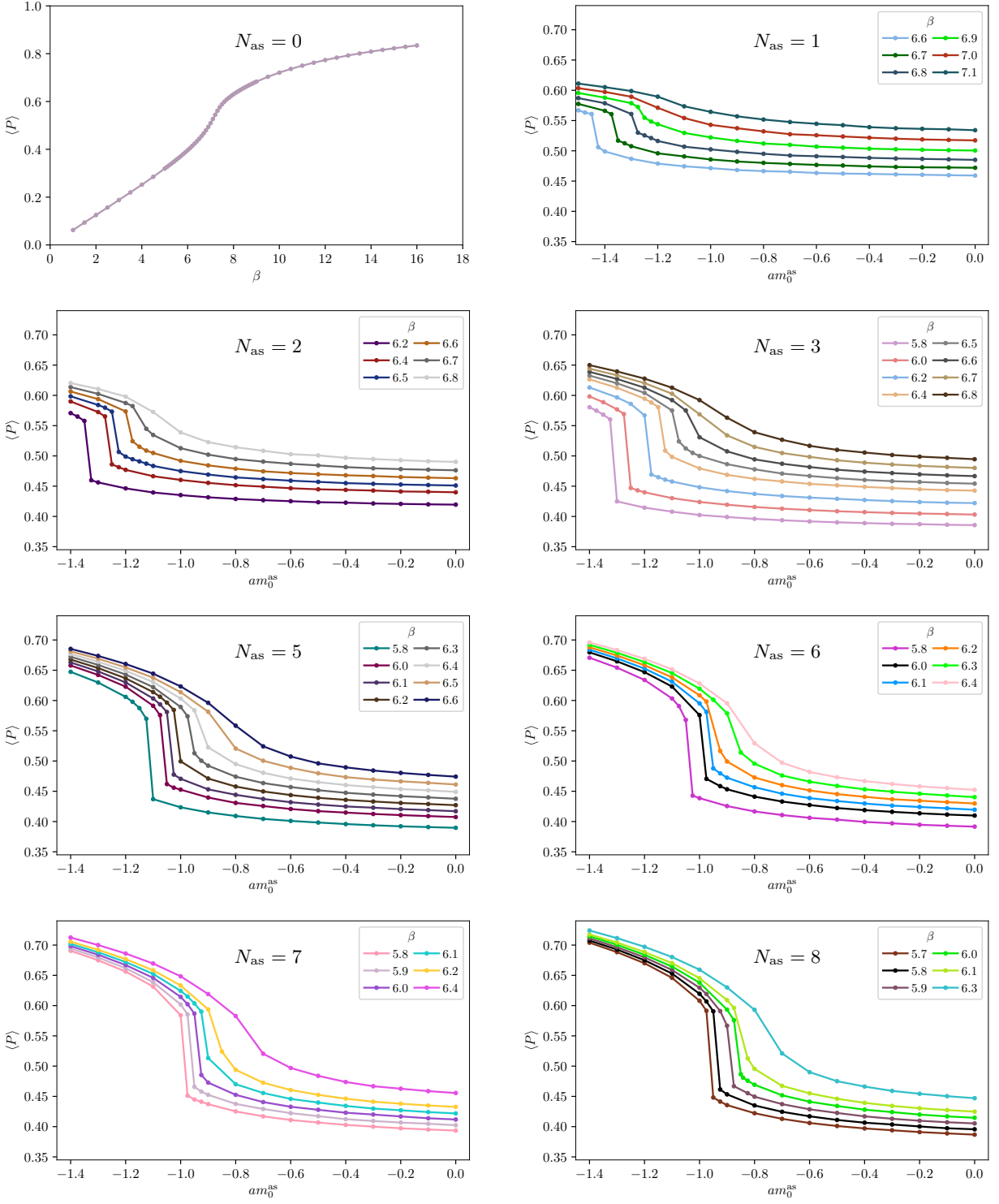

\subsection{Varying $N_{\rm as}$}
\label{Sec:3-5}

We repeat the parameter scan for 
other choices of $N_{\rm as}$,
using the RHMC for all fermions when $N_{\rm as}$ is odd, and the HMC algorithm otherwise.
The purpose of the exercise is to study the dependence of the upper bound coupling for the bulk phase $\beta_{\ast}$ on the number of fermions, $N_{\rm as}$. Indeed, it is expected that for small $N_{\rm as}$ we expect the theory to confine, while for larger values of $N_{\rm as}\sim N_{\rm as}^c$ the theory should approach the lower end of the conformal window,
 and eventually lose asymptotic freedom---we recall that the latter
requires to impose the bound $N_{\rm as}<33/4$ in $Sp(4)$, while setting the stage for a first truly non-perturbative determination
 of the former is the main motivation for this study.

The results of these studies are shown in Fig.~\ref{Fig:3-5}, which displays our measurements of the average plaquette, $\langle P \rangle$, as a function of the bare parameters of the theories.
For the pure gauge $Sp(4)$ theory, we get plaquette values that are in agreement with the ones shown in Ref.~\cite{Holland:2003kg}. The corresponding upper bound value of the coupling is roughly estimated to be $\beta_{\ast} \simeq 7.2$. \\

For theories with dynamical fermions, we vary both the masses and the coupling of the theories. As can be seen from Fig~\ref{Fig:3-5}, for $N_{\rm as}=1$ the upper bound is $\beta_{\ast} \simeq 6.7$. For $N_{\rm as}=2$ the upper bound is $\beta_{\ast} \simeq 6.7$, and for $N_{\rm as}=3$ it is $\beta_{\ast} \simeq 6.5$, in agreement with the values found in Ref.~\cite{Lee:2018ztv}. At a larger number of fermions species, we obtain progressively smaller values of $\beta$ for the upper bound of the bulk phase $\beta$: for  $N_{\rm as}=5$, we get $\beta_{\ast} \simeq 6.3$. For $N_{\rm as}=6$, the upper bound coupling is $\beta_{\ast} \simeq 6.2$. For $N_{\rm as}=7$, we get $\beta_{\ast} \simeq 6.1\div 6.2$ and for $N_{\rm as}=8$, $\beta_{\ast} \simeq 6.1$. \\
Overall, we notice a trend according to which the more fermion flavors are present in the $Sp(4)$, the smaller the upper bound value of the coupling we find and the bigger is the corresponding critical  bare mass $am_0^{\mathrm{(as)}\ast}$.

\section{Scale setting and topology}
\label{Sec:sst}

We return now to the theory with 
$N=2$, $N_{\rm f}=0$, and  $N_{\rm as}=4$.
We discuss a scale setting procedure that uses the Wilson flow. 
We also monitor the evolution of the topological charge, to show that topological freezing was avoided. We focus the discussion on a few representative examples, although we checked that our conclusions have general validity for all choices of parameter relevant to this study.

\begin{figure}[t!]
\begin{center}
\begin{picture}(170,50)
     \put(0,0){ \includegraphics[width=0.44\linewidth]{ 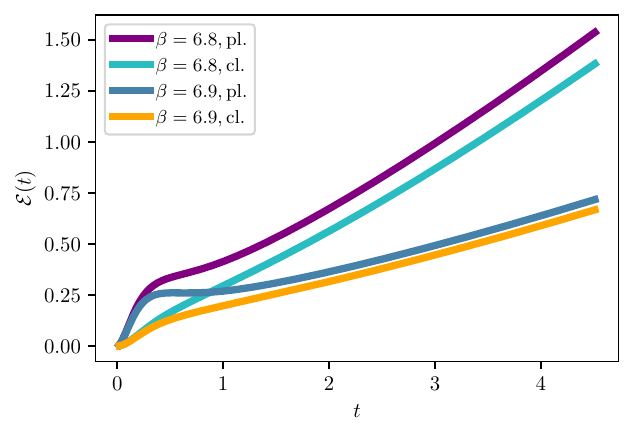}}
     \put(85,0){ \includegraphics[width=0.44\linewidth]{ 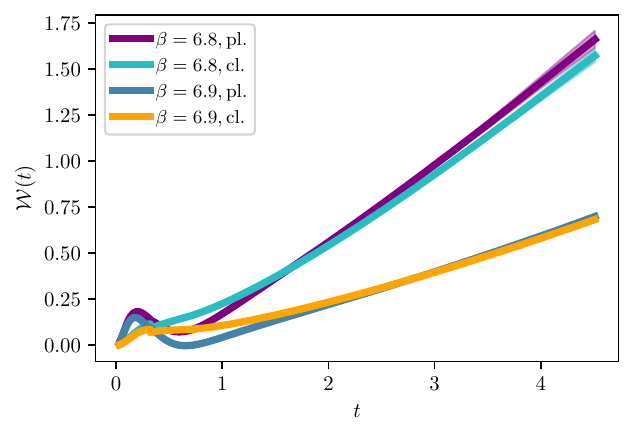}}
      \end{picture}
      \end{center}
  \caption{Wilson Flow~\cite{Luscher:2010iy,Luscher:2013vga} energy density ${\cal E}(t)$ (left panel) and ${\cal W}(t)$ (right),
  computed as in Refs.~\cite{Bennett:2017kga,Bennett:2022ftz}, from the standard (pl) and the clover-leaf (cl) plaquette defined in Refs.~\cite{Sheikholeslami:1985ij,Hasenbusch:2002ai}, for the $Sp(4)$ theory with $N_{\rm as}=4$ fermions transforming in the 2-index antisymmetric representation. The lattice size is $\tilde V = (12a)^4$, and we display two representative choices of bare parameters, with  $\beta = 6.8$ or  $6.9$ and common bare mass $a m_0^{\rm as} = -0.8$. The time step is $0.01$, $t_{max} = 4.5$ to reduce finite-size effects. Errors are computed by bootstrapping. We have chosen ${\cal W}_0 = \frac{1}{2} C_2(F)$ for the topological charge. The corresponding values of $w_0$ from the plaquette and the clover-leaf are $w_{0, pl.} = 1.485(3)$ and $w_{0, cl.} = 1.495(2)$ for $\beta = 6.8$ and $w_{0, pl.} = 2.005(2)$ and $w_{0, cl.} = 2.026(2)$ for $\beta = 6.9$. We have set $a=1$, for notational convenience.}
\label{Fig:WF}
\end{figure}

  \begin{figure}[t]
\begin{center}
\begin{picture}(100,130)
     \put(0,65){ \includegraphics[width=0.50\linewidth]{ 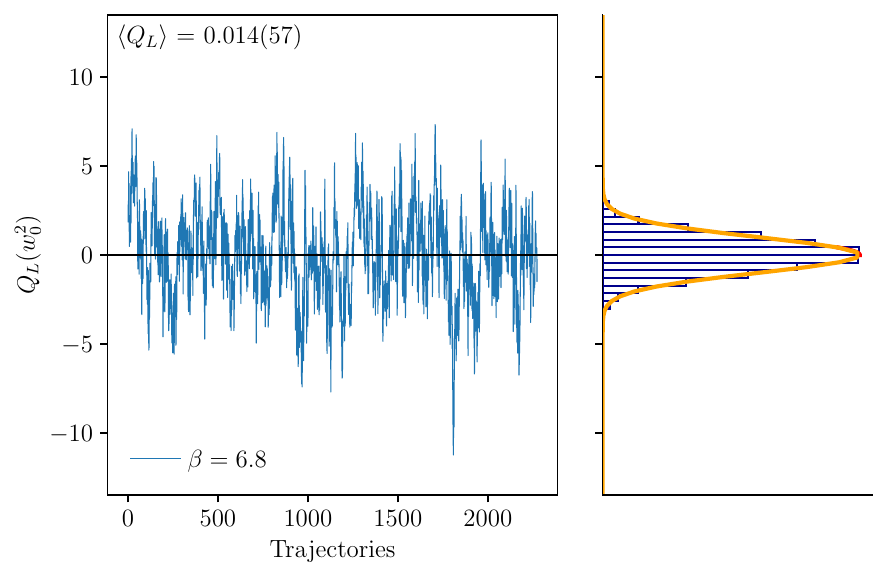}}
      \put(0,0){ \includegraphics[width=0.50\linewidth]{ 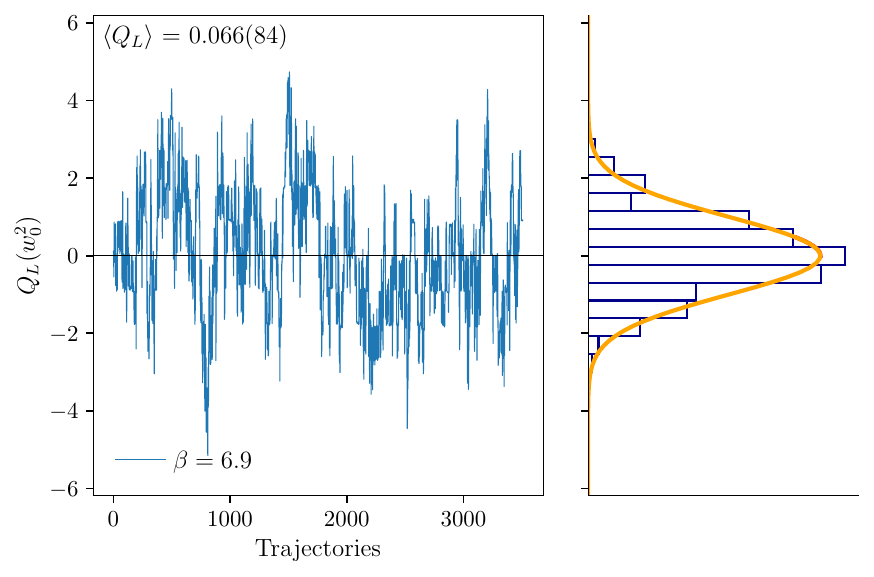}}
      \end{picture}
      \end{center}
   \caption{Evolution with the ensemble trajectories 
   of the  topological charge 
   $Q_L(t=w_0^2) \equiv \sum_x \frac{1}{32\pi^2} \varepsilon^{\mu\nu\rho\sigma} \Tr\left[{\cal C}_{\mu\nu} (x) {\cal C}_{\rho\sigma} (x)\right] $, 
   computed (without rounding) at flow time $t = w_0^2$   for the $Sp(4)$ theory with $N_{\rm as}=4$ fermions transforming in the 2-index antisymmetric representation.  The lattice size is 
   $\tilde V=(12a)^4$. The lattice parameters characterising the ensembles are $\beta = 6.8$ (top panel) and $\beta = 6.9$ (bottom), with bare mass $a m_0^{\rm as} = -0.8$. The histograms of the measurements (right panels) are compatible with a  normal distribution centered at zero, with reduced chi-square $\chi^2 / N_{\rm d.o.f }= \tilde{\chi}^2 = 1.1$ for both panels. The integrated autocorrelation time computed using the Madras-Sokal windowing algorithm is $\tau_Q = 31(3)$ (top)
   and $\tau_Q = 238(12)$ (bottom)}.
 \label{Fig:Q}
\end{figure}

 The gradient flow~\cite{Luscher:2010iy}, and its discretised counterpart, the Wilson flow~\cite{Luscher:2013vga}, 
 are useful for two complementary purposes.
 On the one hand, the Wilson flow provides a universal, well defined way to set the scale in a lattice theory, that is unambiguously defined irrespectively of the properties of the theory and of model-dependent considerations.
 On the other hand, the process we will describe momentarily consists of taking gauge configurations and evolving them with a flow equation, which results in the smoothening of such configurations, and the softening of short-distance fluctuations. The former property is beneficial because it allows to compare to one another different theories for which no experimental information is available (yet), and that might have different matter content. The latter characteristic allows, in practical terms, to reduce the short-distance numerical noise and the effects of discretisation in the lattice calculation of observables, such as the topological charge, $Q$, which are sensitive to fluctuations at all scales. 
 
 We follow Refs.~\cite{Bennett:2017kga,Bennett:2022ftz} (and references therein).
One introduces the flow time, $t$, as an additional, fifth component of the space-time variables, and solves the defining differential equation
 \beqs \label{eq:Wilsonflowdiffeq}
\frac{\mathrm{d} B_\mu(x,\,t)}{\mathrm{d}t} &=&
D_\nu G_{\nu\mu}(x,\,t)\,,
\eeqs
 subject to the boundary conditions $B_\mu(x,\,0)=A_\mu(x)$.
Here $A_{\mu}(x)$ are the gauge fields,
and the covariant derivatives are  
$D_\mu\equiv \partial_\mu + \left[B_\mu,\,\cdot\,\right]$, and
$G_{\mu\nu}(t) = \left[D_\mu,\,D_\nu\right]$.
 As anticipated, the main action of the flow is to introduce a Gaussian  smoothening of the configurations,
 with mean-square radius $\sqrt{8t}$.
 
 In order to use this object to introduce a scale, one defines the quantities
\beqs
 \label{eq:GF_coupling}
    {\cal E}(t)& \equiv& \frac{t^2}{2} \left\langle \mathrm{Tr} \,
    \left[ G_{\mu\nu}(t) G_{\mu\nu}(t)\right]\right\rangle\,,\\
    \label{eq:flowW}
    {\cal W}(t) & \equiv & t \frac{d}{dt}  \mathcal{E}(t) \,,
\eeqs
 and introduces a prescription that defines the scale on the basis of a reference value for either of the two.
 Two common choices in the literature are the scale, $t_0$, defined by setting
 \begin{equation}
\label{eq:scale_t0}
    \left. \mathcal{E}(t)\right|_{t=t_0} = \mathcal{E}_0\,,
\end{equation}
 or the scale, $w_0$, defined implicitly by the condition
\begin{equation}
\label{eq:scale_w0}
   \left. {\cal W}(t)\right|_{t=w_0^2} = \mathcal{W}_0\,.
\end{equation}
Both $ \mathcal{E}_0$ and $\mathcal{W}_0$ are set on the basis of theoretical considerations.
For example, Ref.~\cite{Bennett:2022ftz} advocates to set ${\cal W}_0=c_w C_2(F)$,
where $C_2(F)=(1+2N)/4$ is the quadratic Casimir operator of the fundamental representation in $Sp(2N)$ theories,
and one sets $c_w=0.5$, though other choices are possible.
 
On the discretised lattice, one replaces the gauge field, $A_{\mu}(x)$, with the link variable, $U_{\mu}(x)$,
and the flow equation is rewritten by replacing $B_\mu(x,\,t)$ with the new $V_{\mu}(x,\, t)$ (with $V_{\mu}(x,\, 0) =U_{\mu}(x)$). There are then at least  two ways to replace $G_{\mu\nu}$ with a discretised variable. We introduced the elementary plaquette ${\cal P}_{\mu\nu}$ when defining the 
lattice action in Eq.~(\ref{eq:gauge_action}).
The clover-leaf plaquette operator, ${\cal C}_{\mu\nu}$, provides an alternative to the elementary plaquette,
and can be seen as a simple form of improvement.
We borrow the definition from Refs.~\cite{Sheikholeslami:1985ij,Hasenbusch:2002ai}, that for generic link variables $U_\mu (x)$ reads:
\beqs
\label{Eq:clover-leaf}
\mathcal{C}_{\mu\nu}(x) &\equiv& 
\frac{1}{8}
\left\{ \frac{}{}U_\mu(x) U_\nu(x+\hat{\mu}) U^\dag_\mu(x+\hat{\nu})U_\nu^\dag(x)       +\right.\nonumber\\
        &&\left. 
  + U_\nu(x) U^\dag_\mu(x+\hat{\nu}-\hat{\mu}) U^\dag_\nu(x-\hat{\mu})U_\mu(x-\hat{\mu})\,+\right.\\
        &&\left.
        + \,U^\dag_\mu(x-\hat{\mu}) U^\dag_\nu(x-\hat{\nu}-\hat{\mu}) U_\mu(x-\hat{\nu}-\hat{\mu})
        U_\nu(x-\hat{\nu})  +\right.\nonumber\\
        &&\left.
              + U^\dag_\nu(x-\hat{\nu}) U_\mu(x-\hat{\nu}) U_\nu(x-\hat{\nu}+\hat{\mu})U_\mu^\dag(x)
        -{\rm h.c.}\frac{}{}\right\}\,.\nonumber
        \eeqs

In principle, one would like to set the scale in a way that does not depend crucially on microscopic details. The scale setting using Wilson flow depends on the way the flow equation in Eq.~(\ref{eq:Wilsonflowdiffeq}) is latticised and how the observable $\Tr[G_{\mu \nu}G_{\mu\nu}]$ in Eq.~(\ref{eq:GF_coupling}) is discretised. Therefore, different choices lead to different values for the scale at a given cutoff, but choosing a suitable flow time $t$ allows us to set the scale while reducing drastically these effects.
To this purpose, in Fig.~\ref{Fig:WF} we consider the $Sp(4)$ theory  with $N_{\rm f}=0$ and $N_{\rm as}=4$,
for two representative choices of $\beta$, and a representative choice of volume, $\tilde V$, and bare mass, $a m_0^{\rm as}$,
and we show ${\cal E}(t)$ and ${\cal W}(t)$ as functions of the flow time, $t$, by comparing explicitly the results
obtained by adopting either the elementary or the clover-leaf plaquette as defining the lattice regularisation of the action.
The plots illustrate the general trend  evidenced elsewhere in the literature, according to which the function ${\cal W}(t)$
displays a milder dependence on the short distance regulator. In the following, we set the scale $w_0$ by 
conventionally setting ${\cal W}_0=\frac{1}{2}C_2(F)$.
Recently, several studies for the usage of the Wilson flow observables were performed to define a non-perturbative running coupling--see, e.g., Refs.~\cite{Fodor:2012td,DallaBrida:2016kgh,Hasenfratz:2022zsa,DelDebbio:2021ryq,Fritzsch:2013je} and \cite{DallaBrida:2020pag} for a review. This is a very intriguing application of these tools, but it requires a dedicated study with non-negligible effort and large lattices, which exceed the lattice sizes we explored in this introductory paper.

The topological charge density is defined as
\beqs
q_L(x,t)&\equiv&\frac{1}{32 \pi^2} \varepsilon^{\mu\nu\rho\sigma}\Tr\,\left[{\cal C}_{\mu\nu}(x,t){\cal C}_{\rho\sigma}(x,t)\right]\,,
\eeqs
and the topological charge is $Q_L(t)\equiv\sum_xq_L(x,t)$,
where, again, $t$ is the flow time. In general, the topological charge on the lattice is not quantised,
and in cases where it is the physical quantity of interest---for example because one is working towards a determination of the
topological susceptibility, as in Ref.~\cite{Bennett:2022ftz} and references therein---one needs to evolve to large $t$, and introduce 
a rounding process. 

For the current purposes, we do not need a discretisation algorithm: what we want to verify is that there is no evidence of topological freezing, and to this purpose we perform three simple tests. In Fig.~\ref{Fig:Q} we display the value of $Q_L(t=w_0^2)$
in the $Sp(4)$ theory coupled to $N_{\rm f}=0$ and $N_{\rm as}=4$ fermion species, for two values of the coupling,  $\beta$, 
and a common value of the bare mass. We show how the topological charge evolves along the trajectories,
and supplement it with a histogram displaying its distribution.
Both visual tests confirm that there is no evidence of topological freezing. 
We can make these tests more quantitative by applying the standard Madras-Sokal windowing algorithm~\cite{Madras:1988ei}, and provide estimates of the integrated autocorrelation time $\tau_Q$ of the topological charge, which in both examples, as shown in Fig.~\ref{Fig:WF}, turns out to be many orders of magnitude smaller than the number of trajectories. Furthermore, fits of the histograms are compatible with a Gaussian distribution centered at $\langle Q_L(t=w_0^2)\rangle=0$.

The main message from this section is that the behaviour of the Wilson flow and of the topological charge, computed using the
new software based on Grid, and tested on GPU architecture machines, to examine the properties of the
lattice $Sp(2N)$ gauge theory with $N=2$, $N_{\rm f}=0$, and $N_{\rm as}=4$, provide results that are broadly comparable to
those in the literature for related, though different, field theories. This suggests that the implementation of the simulation routines and of the observables are both free from unwanted effects.

\section{Summary and outlook}
\label{Sec:conclusions}

A number of new physics models based upon $Sp(2N)$ gauge theories has been proposed
in the literature, in such diverse contexts to include Composite Higgs Models, top partial compositeness,
dilaton-Higgs models, strongly interacting dark matter models, among others.
It is essential to the development of all these new physics ideas to provide model builders and phenomenologists 
with non-trivial 
 information about the non-perturbative dynamics.

The programme of systematic characterisation
of $Sp(2N)$ theories
is still in its early stages, though.
Prominently, the challenging question of identifying
the lower end of the conformal window in these theories 
coupled to matter fields in various  representations of the group requires
 the non-perturbative instruments of lattice field theory.
As a necessary step in this direction,
we developed and tested new software, embedded into the Grid environment
to take full advantage of its flexibility.
In this paper we reported the (positive) results of our tests of the algorithms,
that set the stage for future large-scale dedicated studies.
We focused particularly on the $Sp(4)$ theory coupled to $N_{\rm as}=4$  (Dirac) fermions transforming in the
antisymmetric representation, that might be close to the onset of conformality.

We performed a long list of non-trivial exercises. We both tested the effectiveness of the algorithm and software implementation,
but also provided a first characterisation of lattice theories that had never been studied before---although for present purposes we used comparatively small and coarse lattices. We reported in this paper 
 illustrative examples demonstrating that 
there are no obvious problems in the software implementation.
We  computed effectively such observables as the averages of the plaquette and (real) Polyakov loop,
the plaquette susceptibility,  the Wilson flow, and the topological charge.
We  catalogued the first measurements of the critical couplings in $Sp(4)$ lattice theories with
 $N_{\rm as}<33/4$---below the bound imposed  by asymptotic freedom---hence identifying the portion of 
 lattice parameter space connected with the continuum theories of interest.

This paper, and the software we developed for it, set the stage needed
 to explore and quantify the extent of the conformal window in these theories.
The tools we developed can be used also in the context of the
 recent literature discussing the spectroscopy of $Sp(2N)$ theories
with various representations~\cite{Bennett:2017kga,Lee:2018ztv,Bennett:2019jzz,Bennett:2019cxd,Bennett:2020hqd,Bennett:2020qtj,Lucini:2021xke,Bennett:2021mbw,Bennett:2022yfa,Bennett:2022gdz,Bennett:2022ftz,AS,Lee:2022elf,Hsiao:2022kxf,Bennett:2023rsl,Maas:2021gbf,Zierler:2021cfa,Kulkarni:2022bvh,Bennett:2023wjw}, in broad regions of their parameter space, considering both bosonic  bound states as well as fermionic ones, relevant for example in $Sp(2N)$ theories with mixed representations.
This effort can be complemented and further extended by applying  new techniques based on 
the spectral densities~\cite{Hansen:2019idp}---see also the applications in Refs.~\cite{
Hansen:2017mnd,Bulava:2019kbi,Bailas:2020qmv,Gambino:2020crt,Bruno:2020kyl,Lupo:2021nzv,
Gambino:2022dvu,DelDebbio:2022qgu,Bulava:2021fre,Lupo:2022nuj}.
One can envision many more uses and applications of this powerful and flexible open-source software.


\begin{acknowledgments}

The work of EB, JL and BL has been funded by the ExaTEPP project EP/X017168/1. The work of EB and JL has also been supported by the UKRI Science and Technology Facilities Council (STFC) Research Software Engineering Fellowship EP/V052489/1.

The work of NF has been supported by the STFC Consolidated Grant No. 
ST/X508834/1.

The work of PB was supported in part by US DOE Contract DESC0012704(BNL), and in part by the Scientific Discovery through Advanced Computing (SciDAC) program LAB 22-2580.

The work of DKH was supported by Basic Science Research Program through the National Research Foundation of Korea (NRF) funded by the Ministry of Education (NRF-2017R1D1A1B06033701).

The work of LDD and AL was supported by the ExaTEPP project EP/X01696X/1.

The work of JWL was supported in part by the National Research Foundation of Korea (NRF) grant funded 
by the Korea government(MSIT) (NRF-2018R1C1B3001379) and by IBS under the project code, IBS-R018-D1. 

The work of DKH and JWL was further supported by the National Research Foundation of Korea (NRF) grant funded by the Korea government (MSIT) (2021R1A4A5031460).

The work of CJDL is supported by the Taiwanese NSTC grant 109-2112-M-009-006-MY3.

DV is supported by a STFC new applicant scheme grant.

The work of BL and MP has been supported in part by the STFC 
Consolidated Grants No. ST/P00055X/1 and No. ST/T000813/1.
 BL, MP, AL and LDD received funding from
the European Research Council (ERC) under the European
Union’s Horizon 2020 research and innovation program
under Grant Agreement No.~813942. 
The work of BL is further supported in part 
by the EPSRC ExCALIBUR programme ExaTEPP (project EP/X017168/1), by the Royal Society Wolfson Research Merit Award 
WM170010 and by the Leverhulme Trust Research Fellowship No. RF-2020-4619.
LDD is supported by the UK Science and Technology Facility Council (STFC) grant ST/P000630/1.

Numerical simulations have been performed on the Swansea SUNBIRD cluster
(part of the Supercomputing Wales project)
and AccelerateAI A100 GPU system,
and on the DiRAC Extreme Scaling service at the University of Edinburgh.
Supercomputing Wales and AccelerateAI are part funded by
the European Regional Development Fund (ERDF) via Welsh Government.
The DiRAC Extreme Scaling service is operated by
the Edinburgh Parallel Computing Centre
on behalf of the STFC DiRAC HPC Facility (www.dirac.ac.uk).
This equipment was funded by BEIS capital funding
via STFC capital grant ST/R00238X/1 and STFC DiRAC Operations grant ST/R001006/1.
DiRAC is part of the National e-Infrastructure.

\vspace{1.0cm}

{\bf Research Data Access Statement}---The data generated for this manuscript can be downloaded from  Ref.~\cite{12} and the analysis code from Ref~\cite{13}.

\vspace{1.0cm}

{\bf Open Access Statement}---For the purpose of open access, the authors have applied a Creative Commons 
Attribution (CC BY) licence  to any Author Accepted Manuscript version arising.

\end{acknowledgments}

\appendix
\section{Group-theoretical definitions}\label{app:grouptheory}

We denote as $Sp(2N)$ the subgroup of $SU(2N)$ preserving the norm induced by the antisymmetric matrix $\Omega$,
\begin{equation}\label{eq:Omega_matrix}
    \Omega = \begin{pmatrix} 0 & \mathbb{1}_N \\ -\mathbb{1}_N & 0 \end{pmatrix} \;,
\end{equation}
where $\mathbb{1}_N$ is the $N\times N$ identity matrix. This definition can be converted into a constraint on the group element $U$
\begin{equation}\label{eq:UTomegaU_is_U}
    U \Omega U^T = \Omega \; .
\end{equation}
Due to unitarity, the previous condition can be also written as
\begin{equation}\label{eq:UOmega_equals_OmegaUstar}
    U \Omega = \Omega U^* \; ,
\end{equation}
which implies the following block structure
\begin{equation}\label{eq:group_block_structure}
    U = \begin{pmatrix} A & B \\ -B^* & A^* \end{pmatrix} \; ,
\end{equation}
where Eq.~\eqref{eq:UTomegaU_is_U} implies, for $A$ and $B$, that
\begin{equation}
    AB^T = B A^T \; , \;\;\;\;\; A A^\dagger + B B^\dagger = \mathbb{1}_N \; .
\end{equation}
The algebra can be defined by expanding $U \Omega = \Omega U^* $ in terms of the hermitian generators $t^a$, i.e. $U=\exp (i \omega^a t^a)$ for real parameters $\omega^a$. We arrive at the following condition on the generic element of the algebra $T =  \sum_a \omega^a t^a$
\begin{equation}\label{eq:hermitian_algebra_condition_sp2n}
    T \Omega = - \Omega T^* \; ,
\end{equation}
which also implies that 
\begin{equation}\label{eq:algebra_block_structure}
    T = \begin{pmatrix} X & Y \\ Y^* & -X^*  \end{pmatrix} \; .
\end{equation}
Hermiticity imposes the conditions $X=X^\dagger$ and $Y = Y^T$. 
The number of independent degrees of freedom is then $2N(N+1)$, the dimension of the group. 

\section{Generators of the algebra in Grid}\label{app:generators}
Let $t^a_{\rm f}$ be the generators of the Lie Algebra of $Sp(2N)$ in the fundamental representation. They are implemented in Grid as hermitian, meaning that they follow the block structure of Eq.~\eqref{eq:algebra_block_structure}. Their normalisation is such that
\begin{equation}
    \mathrm{Tr} \left( t_{\rm f}^a t_{\rm f}^b \right) = \frac{\delta^{ab}}{2} \; .
\end{equation}
The generators $t^a_{\rm f}$, with $a=1,\dots, \, 2N^2 + N$, are implemented in Grid according to the following scheme. The $2N^2$ off-diagonal generators are identified by the following six relations among their matrix elements:
\begin{equation}\label{eq:lie_generator_atype}
    t^a_{i,j} = t^a_{j,i} = -t^a_{i+N,j+N} = -t^a_{j+N,i+N} = \frac{1}{2\sqrt{2}} \; , \;\;\;\;\; i=1, \dots  N-1 \; , \;\;\;\;\; i<j\leq N \; , 
\end{equation}
with $a = 1 \dots N(N-1)/2$,
\begin{equation}\label{eq:lie_generator_btype}
    t^a_{i,j} = -t^a_{j,i} = t^a_{i+N,j+N} = -t^a_{j+N,i+N} = \frac{i}{2\sqrt{2}} \; , \;\;\;\;\; i=1, \dots, N-1 \; , \;\;\;\;\; i<j\leq N \; ,
\end{equation}
with $a = N(N-1)/2 + 1  \dots N(N-1)$,
\begin{equation}\label{eq:lie_generator_ctype}
    t^a_{i,j+N} = t^a_{j,i+N} = t^a_{i+N,j} = t^a_{j+N,i} = \frac{1}{2\sqrt{2}} \; , \;\;\;\;\; i=1, \dots, N-1 , \;\;\;\;\; i<j \leq N-1 \; ,
\end{equation}
with $a = N(N-1) + 1 \dots 3N(N-1)/2$,
\begin{equation}\label{eq:lie_generator_dtype}
    t^a_{i,j+N} = t^a_{j,i+N} = -t^a_{i+N,j} = -t^a_{j+N,i} = \frac{i}{2\sqrt{2}} \; , \;\;\;\;\; i=1, \dots, N-1 \; , \;\;\;\;\; i<j \leq N-1 \; ,
\end{equation}
with $a = 3N(N-1)/2 +1 \dots 2N(N-1)$
\begin{equation}\label{eq:lie_generators_wtype}
    t^a_{i,i+N} = t^a_{i+N,i} = \frac{1}{2} \; , \;\;\;\;\; i=1, \dots, N \; ,
\end{equation}
with $a = 2N^2-2N+1 \, \dots, \,2N^2-N$,
\begin{equation}\label{eq:lie_generators_ztype}
    t^a_{i,i+N} = -t^a_{i+N,i} = \frac{i}{2} \; , \;\;\;\;\; i=1, \dots, N \; ,
\end{equation}
with $a = 2N^2-N+1, \, \dots, \,2N^2$. The remaining $N$ generators in the Cartan subalgebra are 
\begin{equation}
     (t^a)_{i,i} = -(t^a)_{i+N,i+N} = \frac{1}{2} \; , \;\;\;\;\; i=1, \dots  N \; ,
\end{equation}
with $a = 2N^2 +1 \dots 2N^2 + N$, the dimension of the group. It is useful to provide an explicit representation for $2N=4$:
\begin{equation}
    \begin{split}
        & t_{\rm f}^1 = \frac{1}{2\sqrt{2}} \begin{pmatrix}
            0 & 1 & 0 & 0 \\
            1 & 0 & 0 & 0 \\
            0 & 0 & 0 & -1 \\
            0 & 0 & -1 & 0 \\
        \end{pmatrix} \;\;\;\;\; 
        t_{\rm f}^6 = \frac{1}{2} \begin{pmatrix}
            0 & 0 & 0 & 0 \\
            0 & 0 & 0 & 1 \\
            0 & 0 & 0 & 0 \\
            0 & 1 & 0 & 0 \\
        \end{pmatrix}
        \\
        & t_{\rm f}^2 = \frac{1}{2\sqrt{2}} \begin{pmatrix}
            0 & i & 0 & 0 \\
            -i & 0 & 0 & 0 \\
            0 & 0 & 0 & i \\
            0 & 0 & -i & 0 \\
        \end{pmatrix} \;\;\;\;\;
        t_{\rm f}^7 = \frac{1}{2} \begin{pmatrix}
            0 & 0 & i & 0 \\
            0 & 0 & 0 & 0 \\
            -i & 0 & 0 & 0 \\
            0 & 0 & 0 & 0 \\
        \end{pmatrix}
        \\
        & t_{\rm f}^3 = \frac{1}{2\sqrt{2}} \begin{pmatrix}
            0 & 0 & 0 & 1 \\
            0 & 0 & 1 & 0 \\
            0 & 1 & 0 & 0 \\
            1 & 0 & 0 & 0 \\
        \end{pmatrix} \;\;\;\;\;\;\;\;\;\;
        t_{\rm f}^8 = \frac{1}{2} \begin{pmatrix}
            0 & 0 & 0 & 0 \\
            0 & 0 & 0 & i \\
            0 & 0 & 0 & 0 \\
            0 & -i & 0 & 0 \\
        \end{pmatrix}
        \\
        & t_{\rm f}^4 = \frac{1}{2\sqrt{2}} \begin{pmatrix}
            0 & 0 & 0 & i \\
            0 & 0 & i & 0 \\
            0 & -i & 0 & 0 \\
            -i & 0 & 0 & 0 \\
        \end{pmatrix} \;\;\;\;\;
        t_{\rm f}^9 = \frac{1}{2} \begin{pmatrix}
            1 & 0 & 0 & 0 \\
            0 & 0 & 0 & 0 \\
            0 & 0 & -1 & 0 \\
            0 & 0 & 0 & 0 \\
        \end{pmatrix}
        \\
            & t_{\rm f}^5 = \frac{1}{2} \begin{pmatrix}
            0 & 0 & 1 & 0 \\
            0 & 0 & 0 & 0 \\
            1 & 0 & 0 & 0 \\
            0 & 0 & 0 & 0 \\
        \end{pmatrix} \;\;\;\;\;\;\;\;\;\;\;\;\;\;
        t_{\rm f}^{10} = \frac{1}{2} \begin{pmatrix}
            0 & 0 & 0 & 0 \\
            0 & 1 & 0 & 0 \\
            0 & 0 & 0 & 0 \\
            0 & 0 & 0 & -1 \\
        \end{pmatrix} \; .
    \end{split}
\end{equation}


\end{document}